# StarDICE IV: correcting visible photometry from atmospheric grey extinction using thermal infrared observations

K. Sommer [1,2★] B. Plez [1] J. Cohen-Tanugi [3] M. Betoule [4] S. Bongard [4] T. Souverin,[5] S. Dagoret-Campagne [6] M. Moniez [6] J. Neveu [4,6] F. Feinstein [7] C. Juramy [4] L. Le Guillou [4] E. Sepulveda[4] and E. Nuss[1]†

[1]*Laboratoire Univers et Particules de Montpellier, Université de Montpellier, CNRS, F-34095 Montpellier, France*
[2]*University of Arizona, Steward Observatory, 933 N. Cherry Ave., Tucson, AZ 85721, USA*
[3]*Laboratoire de Physique de Clermont, Université Clermont Auvergne, CNRS, F-63000 Clermont-Ferrand, France*
[4]*LPNHE, CNRS/IN2P3 & Sorbonne Université, 4 place Jussieu, F-75005 Paris, France*
[5]*LAPP, Université Savoie Mont Blanc, CNRS/IN2P3, F-74941 Annecy, France*
[6]*IJCLab, Université Paris-Saclay, CNRS/IN2P3, IJCLab, F-91405 Orsay, France*
[7]*Aix-Marseille Univ., Centre National de la Recherche Scientifique, CPPM, 163 Avenue de Luminy, F-13009 Marseille, France*

Accepted 2026 February 3. Received 2026 January 14; in original form 2025 November 21

## ABSTRACT

Ground-based surveys such as the Vera C. Rubin Observatory's Legacy Survey of Space and Time require photometric calibration that is both long-term stable and spatially uniform at the sub-per cent level, even during non-photometric conditions. Achieving this precision motivates new approaches to characterize atmospheric transmission, particularly to mitigate grey extinction from clouds. The StarDICE experiment aims to establish a metrology chain linking laboratory standards to astrophysical fluxes with 1 mmag accuracy in the *griz* bands, a goal for which controlling variable atmospheric effects is essential. We present a method that corrects photometric measurements using simultaneous radiometric information from an infrared thermal camera. The grey-extinction model is fit on an image-by-image basis using thermal radiance excess and the difference between synthetic and instrumental fluxes of calibration stars, without requiring assumptions about the spatial structure of extinction. The method relies on a forward model that incorporates environmental monitoring, radiative-transfer simulations, and *Gaia* DR3 stellar catalogues. Using data from a remote observing system that repeatedly monitored two fields under diverse atmospheric conditions, we show that the corrections reduce residuals between corrected and reference magnitudes and produce extinction maps with 2-arcmin resolution and ∼0.01 mag accuracy. Using this technique, we can recover data acquired under non-photometric conditions with a precision comparable to data obtained under photometric conditions. For the most affected exposures, the mean absolute error improves from 0.64 to 0.11 mag, and temporal extinction variations can be reduced to 0.025 mag per source. We discuss the implications of this technique for future surveys and outline directions for further refinement.

## 1 INTRODUCTION

The StarDICE experiment (T. Souverin et al. 2024), located at Observatoire de Haute Provence (OHP, France; IAU 511) aims at improving the photometric calibration of CALSPEC standard stars (R. C. Bohlin, K. D. Gordon & P. E. Tremblay 2014; R. C. Bohlin, I. Hubeny & T. Rauch 2020) to the mmag level, by monitoring instrument throughput with a LED-based artificial star source, calibrated on NIST photodiodes (T. C. Larason & J. M. Houston 2008). To project this instrumental calibration to top of the atmosphere spectrophotometric standard stars, StarDICE needs to precisely measure the atmospheric transmission and can only do so in stable photometric condition. One critical point identified in the StarDICE pathfinder experiment (F. Hazenberg 2019) is that even small amounts of extinction variations caused by passing undetected clouds, alters the measurement and drastically reduces the usable statistics. In this specific case standard mitigation procedures which consists in choosing photometric nights where (1) measured flux dispersion is minimal, (2) the atmospheric extinction is somewhat uniform across the field-of-view (FOV) and (3) no visible clouds are present, are not practical (T. Souverin et al. 2024). Another important application of extinction correction techniques are time-domain astronomy surveys such as the Vera C. Rubin Observatory, whose observations will be conducted even under non-photometric condition in order to maximize the telescope duty cycle (D. L. Burke et al. 2013).

★ E-mail: kelian.sommer@lupm.in2p3.fr
† This co-author passed away before the publication of this work.

Given the large FOV (about 9.6 $deg^2$) of the Rubin LSST camera, it is unrealistic to assume that cloud structures are uniform across the entire FOV. The key issue is no longer whether the night is photometric, but rather determining the extent and structure of cloud(s).

The second way consists in exploring methods to correct for cloud extinction by fitting residuals in source photometry. S. Wang et al. (2012) analysed data from the Chinese Small Telescope ARray (CSTAR; see X. Yuan et al. 2008) by constructing a 2D interpolation map of grey extinction using a modified inverse distance weighting algorithm (D. Shepard 1968). They found that over 80 per cent of images were taken under even photometric conditions (extinction unevenness $< 0.01$ mag), while about 8 per cent showed significant extinction variations ($>0.02$ mag). The method was effective in reducing dispersion in non-photometric sequences but showed no improvement under good conditions, and its global impact was not assessed. Ž. Ivezić et al. (2007) calibrated photometric zero points for non-photometric SDSS (D. G. York et al. 2000) data using dense stellar fields, achieving 2 per cent accuracy even through 1 mag thick clouds. They suggested that further improvements could be obtained by modelling cloud opacity across multiple calibration patches, projecting that LSST could reach 1 per cent calibration accuracy through 1–3 mag clouds. Building on this, D. L. Burke et al. (2013) applied 2D polynomial fits to MOSAIC II data and achieved $\sim$0.5 per cent precision in clear conditions and $\sim$1 per cent repeatability through 1.5 mag thick clouds. They also quantified the spatial structure of cloud absorption, noting that sub-per cent precision is limited by the presence of thin cloud layers. Finally, D. L. Burke et al. (2017) introduced the Forward Global Calibration Method (FGCM) for the Dark Energy Survey (Dark Energy Survey Collaboration 2016), obtaining 6–7 mmag residual calibration errors per exposure. Although grey extinction from clouds was not explicitly modelled due to mostly clear observing conditions, FGCM estimated grey extinction as residuals between predicted and observed magnitudes, averaged over each CCD. More recently, S. Garrappa et al. (2025) introduced a method of absolute calibration that simultaneously treats instrumental and atmospheric effects on an image-by-image basis by fitting the system transmission. They proceed in a similar manner to D. L. Burke et al. (2013) by fitting zero-point distribution using 2D polynomials on sub-divisions of each single image and obtain an accuracy of the zero-point between 3 and 5 mmag for each exposure when applied on data from the Large Array Survey Telescope (LAST, see E. O. Ofek et al. 2023). All these methods rely on high-density stellar fields within the instrument's FOV and show diminished performance when stellar density is low or when grey extinction exceeds $\sim$1.5 mag (D. L. Burke et al. 2013).

We investigate here another approach that has been proposed in the conclusions of D. L. Burke et al. (2013) and F. Hazenberg (2019) and which consists in gathering independent measurements of atmospheric down-welling radiance using an infrared thermal camera. Our goal is to maintain sufficient precision on individual observations collected under non-photometric conditions while maximizing the StarDICE telescope uptime. We present our method to correct optical photometry from cloud extinction using thermal infrared observations in Section 2. We then present the instrumental setup in Section 3, and detail the observation campaign and collected data in Section 4. Sections 5 and 6 are respectively devoted to the presentation of the analysis and results. Finally, in Section 7, we summarize the main findings of this study, and we discuss current limitations and potential improvements.

## 2 METHODOLOGY

### 2.1 Cloud visible extinction and IR radiance

Thin cirrus clouds can be nearly transparent in the visible spectrum – while still inducing extinction – but nearly opaque across the 8–13 µm window region (W. L. Smith et al. 1998). Indeed, these clouds are composed of water droplets and ice crystals that emit thermally in the long-wave infrared (LWIR) spectrum at a temperature range $T$ of 200 to 300 K (A. J. Heymsfield et al. 2017).

D. H. DeSlover et al. (1999) demonstrated that visible and IR optical cloud depths are related by proportionality when assuming the cloud is composed of ice droplets of uniform size. Furthermore, D. Serrano et al. (2015) showed that the optical thickness of clouds is almost independent of wavelength in the UV to near IR.

Under these hypotheses, the broad-band magnitude excess due to the cloud extinction reads $\Delta m \approx 1.086\,\tau_{\mathrm{VIS}} \approx 1.086\,a\,\tau_{\mathrm{IR}}$, where $\tau_{\mathrm{VIS}}$ and $\tau_{\mathrm{IR}}$ are the cloud optical thickness in the visible and IR region, respectively, and $a$ is their ratio, which depends on the ice crystal size only. The radiation transfer equation applied to the cloud excess radiance $\Delta L$ immediately yields,

$$\Delta L = A \times (1 - \exp(-B \times \Delta m)), \quad (1)$$

where $A$ and $B$ are parameters to be determined from the data ($A$ is directly related to the Planck function at the temperature of the cloud). Infrared thermal cameras are known to be able to observe the sky radiance in the 8–13µm window (e.g. J. A. Shaw & P. W. Nugent 2013), but cloud structures that cause grey extinction vary on time-scales shorter than the typical exposure times of large survey telescopes, and on spatial scales much smaller than their full field of view (A. Fliflet & W. Manheimer 2006; Ž. Ivezić et al. 2007; D. L. Burke et al. 2013). As a consequence, coincident observations in the visible (for $\Delta m$) and LWIR (for $\Delta L$) are warranted, together with a careful modelling of the cloud-free atmosphere in order to derive the excess radiance and the visible extinction from observations.

### 2.2 Forward model for the cloud-free atmosphere

To estimate the radiance excess $\Delta L$ of clouds from radiometric images, it is essential to know and subtract the down-welling radiance of a cloud-free atmosphere. Indeed, the LWIR spectral region contains atmospheric emission lines mostly from water, carbon dioxide, and ozone (see Fig. 1). Water vapour is poorly mixed in the atmosphere and exhibits significant short-term variations (W. M. Wood-Vasey, D. Perrefort & A. D. Baker 2022), which strongly affect the atmospheric thermal infrared radiance. Ozone and carbon dioxide also contribute to the total radiance measured by the instrument. Ozone is uniformly distributed and varies little over the course of a night but does vary seasonally (C. W. Stubbs et al. 2007). The same applies to carbon dioxide, whose concentration fluctuates by around ten parts per million (ppm) daily, according to measurements made at OHP over the period 2014 to 2020 (L. Lelandais et al. 2022). Their contribution to LWIR radiance is therefore easier to calculate with precision, and computation of the infrared radiance – and the optical transmission – from the top-of-atmosphere to the Earth's surface, over a wide variety of conditions, can be achieved with extremely good accuracy with modern, readily available, radiative transfer computer program like LIBRADTRAN (B. Mayer & A. Kylling 2005; C. Emde et al. 2016).[1] The program solves the radiative transfer equation for a given atmospheric setup and outputs multiple physical quantities, including the thermal infrared radiance as a function of wavelength, as well as the atmospheric transmission. We use the DISORT (DIScrete Ordinate Radiative Transfer; see K. Stamnes et al. 1988) solver and the pseudo-spherical atmosphere approximation to accurately account for the effects of atmospheric curvature. The contribution due to solar radiation in the LWIR channel is not considered in the simulations. We use the highest available molecular spectral resolution of 1 $cm^{-1}$ based on data from HITRAN (L. Rothman et al. 2005).

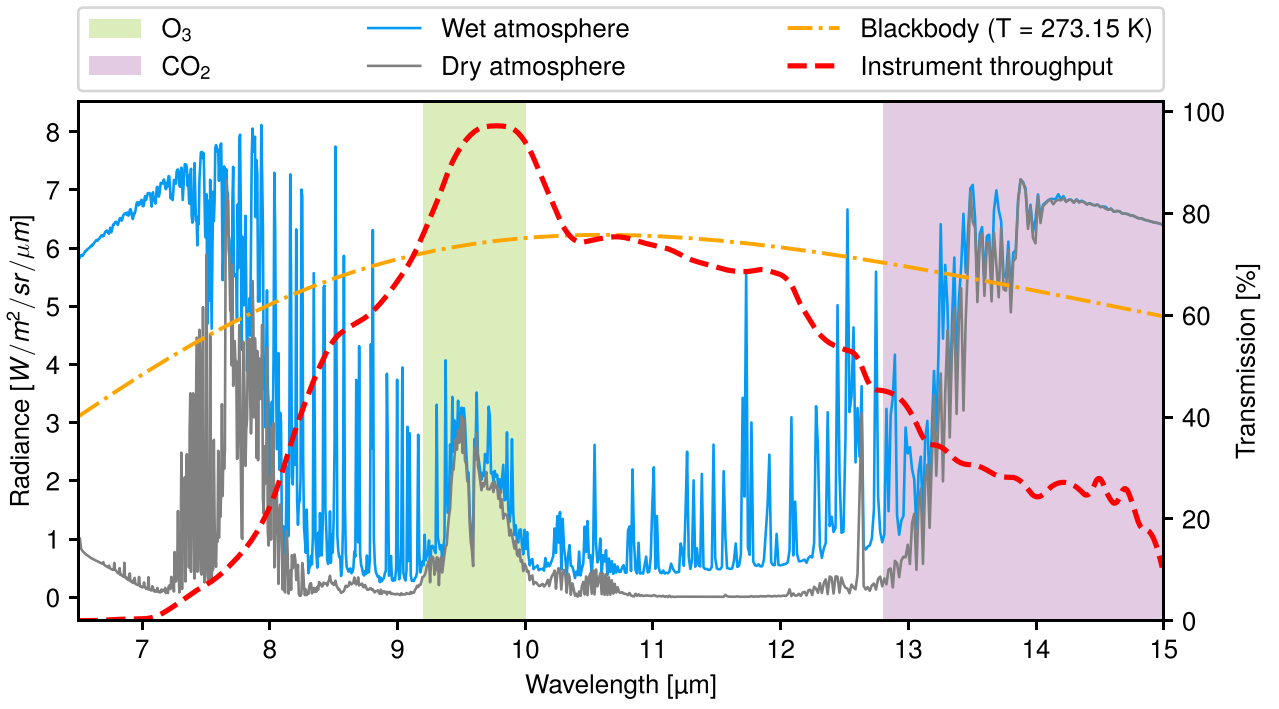

**Figure 1.** Synthetic sky down-welling radiances (solid curves) against instrument transmission (red dashed curve) as a function of wavelength in the long-wave infrared (LWIR) band. The radiances are simulated at airmass $X = 1$ with LIBRADTRAN for a typical atmosphere encountered at OHP site (650 m ASL, $P_{atm} = 937.5$ hPa, $O_3 = 350$ DU, PWV = 12 mm) using a summer seasonal profile. The blue curve depicts the standard atmosphere including the water vapour content whereas the grey curve illustrates the dry component. The orange dash-dotted curve corresponds to the radiance of a blackbody at 273.15 K. The green and purple regions illustrate wavelength ranges dominated by ozone and carbon dioxide emissions, respectively.

### 2.3 Operations and analysis overview

A schematic overview of this work is presented in the flowchart of Fig. 2. Photometric and radiometric observations are simultaneously collected using a CMOS detector attached to a telescope (Section 3.1), and an infrared thermal camera (Section 3.2), respectively, on the same line of sight. In parallel, ancillary instruments (Section 3.3) measure the environmental conditions whose physical quantities are required for Earth's atmosphere numerical simulations.

The forced aperture photometry of astrophysical sources (Section 4.1) follows quality selection cuts (Section 5.1). It is corrected for the chromatic extinction component (Section 5.4) by subtracting the synthetic photometry of the stars (Section 5.2) crossing a simulated clear-sky atmosphere. The synthetic photometry is performed using a data base of stellar spectra, with stellar parameters from the Gaia Data Release 3 catalogue (Gaia Collaboration 2016, 2023). It also requires prior knowledge of the instrumental throughput. Then, the estimation of the grey extinction affecting each star in each exposure is computed as differences between its reference magnitude and its chromatic extinction-corrected instrumental magnitudes (Section 5.5).

[1]Version 2.0.6 from December 24, 2024: https://www.libradtran.org/doku.php

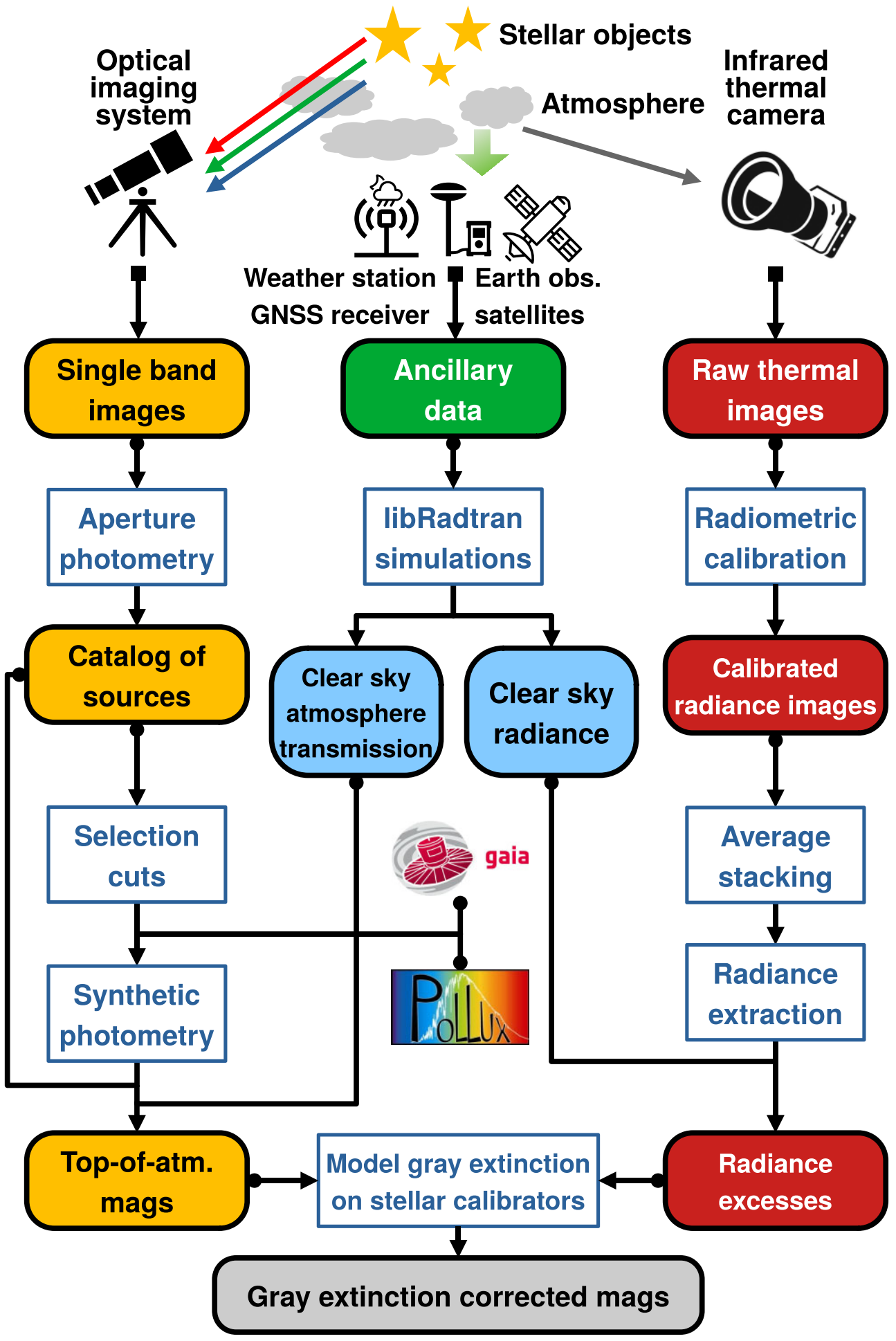

**Figure 2.** Flowchart of the analysis presented in this work. Data products are indicated by rounded boxes, while discrete stages of our analysis are indicated with rectangular boxes.

Raw thermal images undergo radiometric calibration before radiance extraction in the infrared average stacked image at the corresponding location of the stars (Section 4.2). The simulated cloud-free sky down-welling radiance (Section 2.2) is subtracted, to produce radiance excesses.

A fraction of the stellar sources present in the photometric field are used as calibrators to fit the grey extinction model (Section 5.6) expressed as a function of radiance excess. The model is then used to correct the photometry of the objects of interest in the same field (Section 5.7).

## 3 INSTRUMENTAL SETUP

The prototype system for the proof of principle presented in this work is installed at the Observatoire de Haute-Provence. It includes the main imaging systems as well as additional environmental sensing devices.

### 3.1 Optical imaging system

The optical imaging system comprises (1) a Sky-Watcher EQ6-R PRO Go-To equatorial mount (ref. S30300), (2) a TS-Optics apochromatic refractor with 72 mm aperture diameter and f/6 focal ratio (ref. TSAPO72F6), (3) a TS-Optics 0.8× flat-field corrector and focal reducer (ref. TSRed08-72) providing a 346 mm focal length and f/4.8 effective aperture, (4) a ZWO ASI183MM-Pro camera and (5) a Baader *r'* filter (ref. 2961703R), that emulates the *r* passband from the SDSS photometric system. The ZWO ASI183MM monochrome camera features a 20-megapixel back-illuminated Sony IMX183 CMOS 1 arcmin sensor with 2.4 µm pixel size, 12-bit ADC, and low readout noise $\leq 2.5\,e^-$. It is cooled to 0°C to reduce thermal noise during data collection (i.e. dark current noise $\leq 0.006\ e^-\ s^{-1}\ pixel^{-1}$.[2]). The signal-to-noise ratio (SNR) is further improved by using 2 × 2 binning, producing an image with a scale of 2.86 arcsec pixel$^{-1}$, and a field-of-view of 2.18 × 1.46 deg. The total theoretical instrumental spectral throughput curve $\mathcal{T}^{inst}(\lambda)$ computed from manufacturers' individual nominal transmission curves is depicted in Fig. 3.

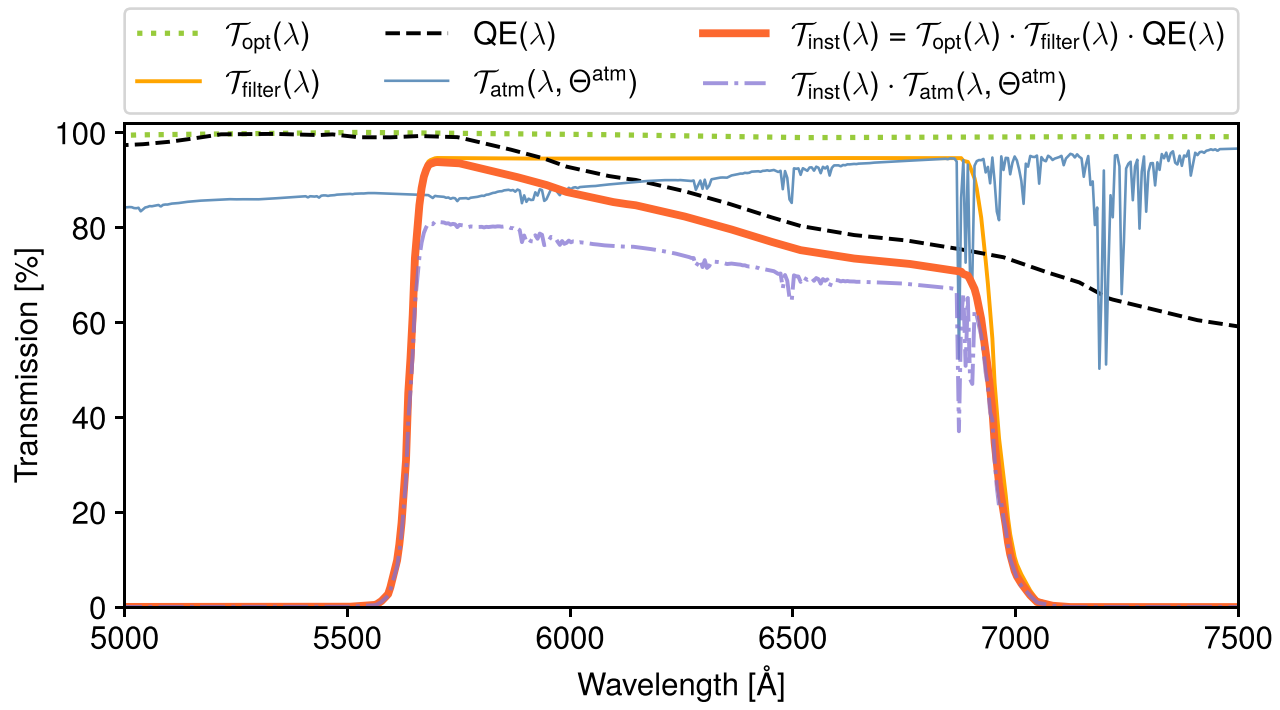

**Figure 3.** Synthetic transmission curves for the elements of the instrumental chain used for photometry. The optical transmission curve (dotted green curve), filter transmission curve (solid orange curve), the detector's quantum efficiency (dashed black curve) and the atmospheric transmission curve (solid blue curve; same parameters as the simulated curve of Fig. 1) are also shown. The solid bold red curve correspond to the theoretical instrumental transmission $\mathcal{T}^{inst}(\lambda)$ without atmosphere defined in equation (11). The dash dotted purple curve corresponds to the product of the instrument and atmosphere transmission curves. The vertical axis represents the global standardized transmission.

### 3.2 Thermal infrared imaging system

Used across a wide range of domains, thermal IR imaging devices have already been successfully exploited for targeted all-sky monitoring applications related to astronomical observations (RASICAM see P. M. Lewis, H. Rogers & R. H. Schindler 2010; K. Reil et al. 2014 at CTIO; SASCAM see J. Sebag et al. 2010, and for cloud monitoring at Vera C. Rubin Observatory; CFHT all-sky camera W. Mahoney, G. Morrison & G. Matsushige 2012). However, their on-sky spatial resolution is only about 20 arcmin, due to their small number of large pixels ($\leq$ 0.5 mega-pixel and 25 µm pixel size for the experiments cited above) and their field of view of about 180°. In order to characterize cloud structure on scales appropriate for optical photometry (i.e. on a star-by-star basis or image portions), we need to resolve smaller scales.

The focal plane array (FPA) of IR thermal cameras is composed of microbolometers (A. Rogalski & K. Chrzanowski 2014). The captured IR radiation produces an electrical signal in the thin metallic film of the sensor, converted to ADU. When uncooled, these cameras lack thermal stabilization, and require specific calibration procedures to account for multiple temperature dependence effects, and to transform the raw ADU to true scene radiance (see O. González-Chávez et al. 2019 for a review of existing methods).

For the LWIR radiometric imaging system, we use a FLIR Tau2[3] thermal infrared camera operating in the 8–14 µm) that has undergone prior radiometric calibration (K. Sommer et al. 2024). The camera core is coupled to a Umicore athermalized 60 mm f/1.25 lens, which provides a scale of 58.4 arcsec pixel$^{-1}$. The sensor is an array of 640 × 512 vanadium oxide (VOx) square pixels of 17µm size. The FOV of an image is 10.37 × 8.30 deg. In radiometric mode, the incident radiance is digitized over a 14-bit depth at a frame rate of 8.33 Hz provided by the ThermalGrabber USB 2.0 interface from TeAx[4] connected to the control microcomputer. A custom Python script[5] is used to send commands over the camera's serial bus. The raw images are then calibrated during post-processing to produce images in physical radiance units (W m$^{-2}$ sr).

To perform *in-situ* non-uniformity correction, an external thermal flat-field support is added. It is made of a thin 2 mm copper plate placed at ∼5 cm from the camera, facing the lens and covered with high-emissivity ($\epsilon = 0.95 \pm 0.01$) 3M Scotch Super 88 Vinyl electrical tape (N. Avdelidis & A. Moropoulou 2003; D. Benirschke & S. Howard 2017). The plate is attached to a high-torque servomotor that rotates it in front of the camera lens between two consecutive photometric exposures.

All components are placed on an aluminum plate and dovetail bar onto the mount head. The IR and optical instruments are approximately aligned manually, which is sufficient as the FOV of the IR camera is much wider than the one from the refractor.

### 3.3 Ancillary equipment

Precise numerical simulations of the atmosphere require several pieces of information about the local meteorological and atmospheric conditions as input parameters. We make use of the StarDICE weather station (Vaisala WTX536[6]), which samples air temperature, relative humidity, ground atmospheric pressure and other parameters, every 10 s, continuously. At each observing epoch, weather readings are saved for post-processing operations (see Section 5). The precision on barometric pressure readings $P_{atm}$ is better than 0.15 hPa at the 1-$\sigma$ level. An accurate sensing of precipitable water vapour (PWV) is also critical for both radiometric measurements (B. Thurairajah & J. Shaw 2005) and photometric observations (W. M. Wood-Vasey et al. 2022). Therefore, we installed a GNSS-based measurement system (see J. Sugiyama, H. Nishino & A. Kusaka 2024, for reference), which consists of a multiband GNSS receiver (u-blox ZED-F9P) coupled with a high-precision antenna (Tallysman VSP6037L1) positioned on the StarDICE observatory roof. This receiver is configured to output raw GNSS data every second, saved in daily files for further processing with Precise Point Positioning (PPP) softwares/services to estimate the PWV above the observatory site at any epoch with sub-millimetre accuracy.

[2]See manufacturer's specifications: https://www.zwoastro.com/product/asi183/

[3]FLIR Systems Inc., Wilsonville, OR, USA
[4]https://thermalcapture.com/tcg/
[5]https://github.com/Kelian98/tau2_thermalcapture
[6]https://www.vaisala.com/en/products/weather-environmental-sensors/weather-transmitter-wxt530-series

## 4 DATA

An observation campaign was conducted remotely over a three-month period, from May to August 2024 at the OHP. Radiometric, photometric, and part of ancillary data were gathered nightly using the setup described in Section 3. Further details on the observation campaign are reported in Appendix A, and the ancillary data is described in Appendix B.

### 4.1 Photometric data processing

#### *4.1.1 Detrending*

A master bias frame was generated by averaging the 25 bias images from the night, and was subtracted from science images. No dark or flat frames were used because (1) the dark current of the CMOS camera is negligible for 20 s exposure time; (2) there is no adequate equipment (e.g. wall-mounted flat screen, flip-flat, or cover) to capture flat frames and the FOV is estimated to be too large to guarantee the uniform illumination across the entire image when doing sky flats.

#### *4.1.2 Astrometry*

The astrometric resolution of each science image was calculated using ASTROMETRY.NET software (D. Lang & others 2010). The World Coordinate System (WCS) solution was stored in the CMOS image FITS file headers. Fitting residuals in both the right ascension and declination axis were typically less than 0.2 arcsec RMS, which is sufficient for source positioning with the Gaia Data Release 3 (DR3) (Gaia Collaboration 2016, 2023), given the imaging angular resolution of 2.87 arcsec pixel$^{-1}$.

This astrometry process is essential to (1) identify the sources in a reference system through a sequence of images and carry out forced photometry, (2) correct the CMOS images from geometric distortion, (3) determine the exact position of each pixel in the infrared field by coordinate transformation between the reference frames, (4) estimate the airmass for each pixel in both the CMOS and infrared field which is required to simulate the chromatic atmosphere transmission and the cloud-free radiance.

#### *4.1.3 Source catalogue*

The catalogue of sources is built by matching observed stars with those from *Gaia* DR3 catalogue whose parameters have been fitted (M. Fouesneau et al. 2023). Using the astrometric solution of a single image from the sequence, we defined a circular region with a 1.75° radius centred on the image. We then searched for all stars with a magnitude of $r \leq 14.5$ from the *Gaia* DR3 catalogue, accessed through the VizieR service (F. Ochsenbein, P. Bauer & J. Marcout 2000) via the ASTROQUERY package (A. Ginsburg et al. 2019). The $r$ magnitude used to apply this cut was computed from the colour transformation[7] from the *Gaia* DR3 photometric bands,

$$\begin{aligned} G - r = &-0.09837 + 0.08592 \times (G_{\mathrm{BP}} - G_{\mathrm{RP}}) \\ &+ 0.1907 \times (G_{\mathrm{BP}} - G_{\mathrm{RP}})^2 \\ &- 0.1701 \times (G_{\mathrm{BP}} - G_{\mathrm{RP}})^3 \\ &+ 0.02263 \times (G_{\mathrm{BP}} - G_{\mathrm{RP}})^4, \end{aligned} \tag{2}$$

where $G$, $G_{\mathrm{BP}}$, and $G_{\mathrm{RP}}$ are the bands defined by the *Gaia* DR3 photometric system (M. Riello et al. 2021). The uncertainty at $1\sigma$ given along this photometric system conversion function is $\sigma_{\mathrm{Gaia}} = 0.037$ mag. In this way, all the stars potentially observed by the telescope will be processed for subsequent photometry, independently of their extinction on a specific image. The advantage of the *Gaia* catalogue is that it also provides stellar parameters (effective temperature $T_{\mathrm{eff}}$, surface gravity $\log g$, metallicity [Fe/H], see M. Fouesneau et al. 2023) that are useful for our analysis, particularly for estimating the impact of the chromatic component of the atmospheric transmission (see Section 5.4).

#### *4.1.4 Aperture photometry and instrumental magnitudes*

We performed forced aperture photometry to extract flux from the sources at the positions listed by the *Gaia* DR3 catalogue.

The right ascension $\alpha$ and declination $\delta$ of each source in the *Gaia* DR3 catalogue were converted into pixel coordinates $x, y$ using the `astropy.skycoord_to_pixel` function and the WCS metadata contained in the headers of the science images. These coordinates correspond to the approximate position of the centre of the star images. However, remaining field distortion – noticeable at the edges of the images even after correction – causes a displacement of a few pixels from the star's centre. Therefore, precise centring of the sources (i.e. centroid determination) was executed using the centre of mass calculation (i.e. the method of moments, see C. Fosu, G. Hein & B. Eissfeller 2004) via the `photutils.centroids` module[8] (L. Bradley et al. 2024).

We integrated the ADU count of each pixel located within an aperture with a radius varying from 1 to 20 pixels using the `sum_circle` function from SEP[9] (E. Bertin & S. Arnouts 1996; K. Barbary 2016). For our case study, subtracting the local background around the star was more appropriate than a global background correction, as it is not homogeneous in non-photometric conditions. The local background level around each source was estimated using a circular annulus with inner and outer radii 3 and 6 pixels larger than the aperture radius.

The instrumental magnitude in the $b$ band[10] for a source $s$ in an image $i$ is determined as follows,

$$m^{\mathrm{inst}}_{b,s,i,k} = -2.5 \times \log_{10}\left(\frac{\mathrm{ADU}_{b,s,i,k}}{t_{\mathrm{exp}}}\right) + \Delta_{b,i,k}, \tag{3}$$

where $\mathrm{ADU}_{b,s,i,k}$ are the background subtracted ADUs collected in the aperture $k$, $t_{\mathrm{exp}}$ is the exposure time in seconds, and $\Delta_{b,i,k}$ is the aperture correction. The reference aperture is not large enough to prevent the measurements from being influenced by seeing variations. Consequently, we applied aperture corrections using the method and model of P. B. Stetson (1990). For each image, an average growth curve was fitted to produce an aperture correction in magnitude $\Delta_{b,i,k}$, where $k$ is the aperture radius.

The photometric uncertainty $\sigma^{\mathrm{phot}}_{s,i}$ was calculated via the `calc_total_error` function from the ASTROPY[11] package (Astropy Collaboration 2022), which determines the total noise in the image based on source (shot noise) and background counts. Then, the function `sum_circle` from SEP package determined the noise in the aperture for each source. The aperture correction uncertainty $\sigma_{i,k}(\Delta_{i,k})$ was calculated from the fitted aperture correction model uncertainty. We selected an aperture radius of $k = 4$ pixels which gave the optimal results considering noise, seeing variations and neighbours contamination. The total photometric uncertainty was then defined as:

$$\sigma^2(m^*_{s,i}) \equiv \left(\sigma^{\text{phot}}_{s,i}\right)^2 + \sigma^2_{s,i,k}(\Delta_{i,k}) \quad . \tag{4}$$

[7] Photometric relationships of Gaia DR3 with other photometric systems: https://gea.esac.esa.int/archive/documentation/GDR3/Data_processing/chap_cu5pho/cu5pho_sec_photSystem/cu5pho_ssec_photRelations.html

[8] Version 2.2.0: https://photutils.readthedocs.io/en/2.2.0/

[9] Version 1.4.1: https://sep.readthedocs.io/en/stable/index.html

[10] We drop the $b$ subscript in the following, as only one bandpass is considered in this study.

[11] Version 7.1.0: https://docs.astropy.org/en/stable/

## 4.2 Radiometric data processing

### 4.2.1 *Calibration*

Raw infrared thermal images are calibrated and converted into radiance units (W m$^{-2}$ sr) following the expression from K. Sommer et al. (2024),

$$L^{\text{inst}}_{\text{obs}} = g_{i,j}\,(S_{i,j} - o_{i,j}) - \alpha_{i,j}\,L^{\text{sens}}_{\text{cam}}(\text{T}_{\text{cam}}) + \beta_{i,j}\,L^{\text{sens}}_{\text{pix}}(\text{T}_{\text{fpa}}), \tag{5}$$

where $S_{i,j}$ is the camera's raw response of the (i, j) pixel expressed in ADUs, $g_{i,j}$, $o_{i,j}$, $\alpha_{i,j}$, and $\beta_{i,j}$ are calibration parameters. $L^{\text{sens}}_{\text{cam}}$ represents the camera housing radiance at temperature $\text{T}_{\text{cam}}$ and $L^{\text{sens}}_{\text{pix}}$ is the microbolometer's radiance at $\text{T}_{\text{fpa}}$.

### 4.2.2 *Radiance extraction and estimation of radiance excess*

Given the spatial resolution of approximately 1 arcmin pixel$^{-1}$ of the IR camera, each stellar source – spanning a few arcseconds – is contained within a single IR pixel. Consequently, the IR radiances used to describe the grey extinction affecting a source are extracted from this pixel.

The geometric transformation between the optical and LWIR field-of-views was achieved by simultaneously observing the Moon with both instruments, as it is the only object emitting sufficient flux for both sensors, stars being invisible to the infrared sensor. Calibration images were captured after slewing the equatorial mount along the right ascension and declination directions. These images enabled the computation of translation, rotation, and scaling matrices that relate the two instruments' celestial footprints. Finally, the pixel-to-celestial coordinate transformation for the LWIR instrument was carried out using the `astropy.wcs.utils.pixel_to_skycoord` function, which converts the LWIR image's pixel coordinates relative to the CMOS pixel coordinate system.

The position of stars fluctuates during the observations (due to the lack of autoguiding, the mount's periodic error caused by the right ascension worm gear, and small polar alignment error), so the extraction is performed using temporal interpolation of the floating position of the source for each optical exposure, via nearest-neighbour interpolation using the `interpolate.RegularGridInterpolator` function from the SCIPY (P. Virtanen et al. 2020) package. In Fig. 4(a), the position of the training and test stars on the CMOS detector field of exposure n°2525 are plotted. It can be seen in Fig. 4(b) that the optical field occupies only a small portion of the total IR field, spanning a rectangle of approximately 131 × 88 pixels. Fig. 4(c) shows the time series of the radiance extracted at the illustrated positions of the three selected stars during the single 20 s concurrent optical exposure. It can be noticed that the high acquisition rate and sensitivity of the IR camera make it possible to capture fast and low variations of the sky down-welling radiance on a pixel-by-pixel basis.

Lastly, once the average radiance for each source $s$ across each image $i$ is extracted, the radiance excess defined in equation (1)

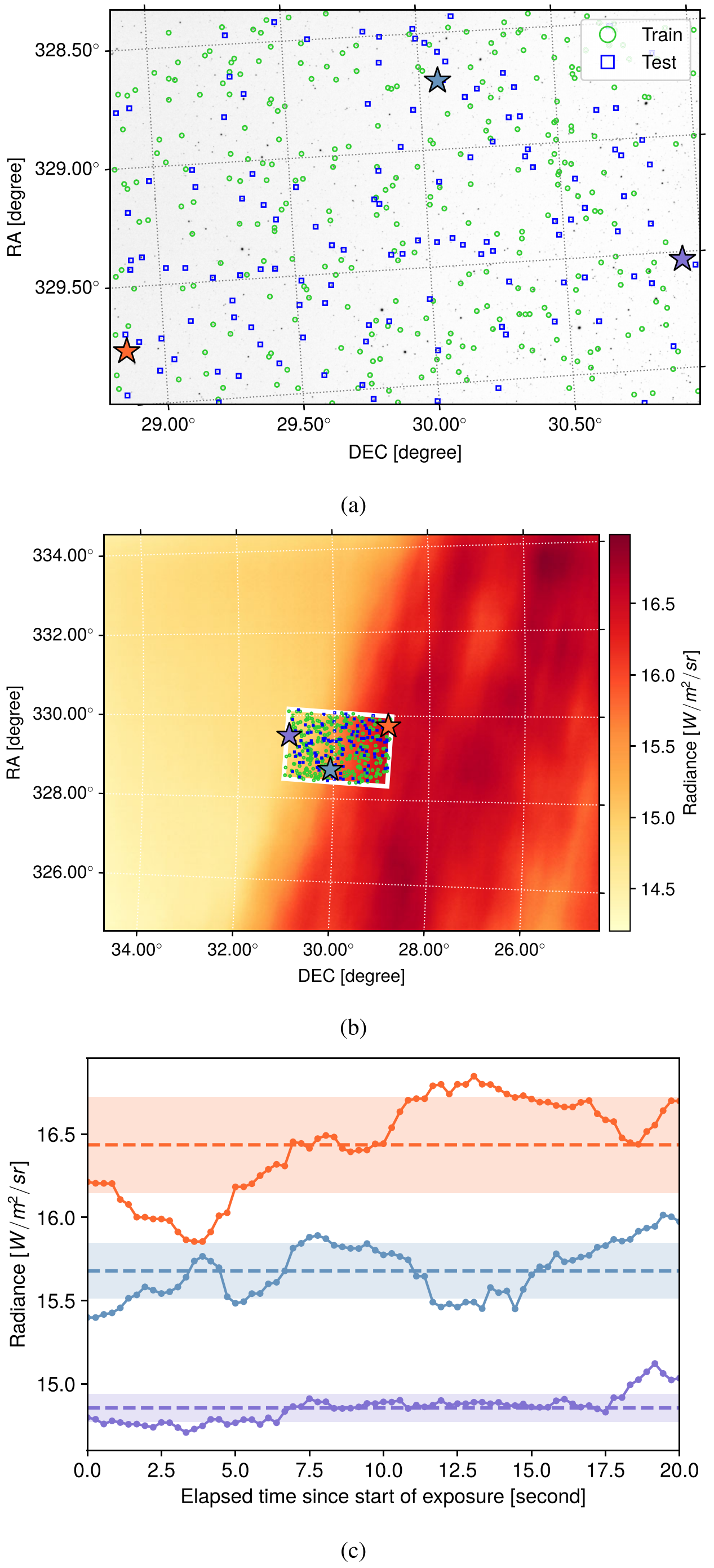

**Figure 4.** Panel (a): The photometric image of the test field near BD+28 4211, centred approximately at R.A. = 329° and DEC. = 29°. The field contains roughly 425 stars after selection cuts. Green circles and blue boxes indicate the positions of training and test sources, respectively. Panel (b): Stacked radiometric image captured simultaneously with the photometric image. The white rectangle delimits the photometric field-of-view overlaid onto the radiometric image. Higher IR radiances due to clouds are shown in red. Panel (c): Time series of radiance for the three colour-matching sources marked above. Radiance values are extracted from 73 IR images acquired during a single 20 s optical exposure. The horizontal dashed lines represent the mean radiance, while the shaded areas indicate the interval of ± 1-$\sigma$ around the mean.

can be calculated as,

$$\Delta L_{s,i} = L_{s,i}^{\rm obs} - L_{s,i}^{\rm sim}, \quad (6)$$

where $L_{s,i}^{\rm sim}$ is the expected at sensor cloud-free radiance simulated with LIBRADTRAN (see Section 2.2).

## 5 ANALYSIS

The analysis proceeds along the following steps:

(i) Forced photometry of sources
(ii) Selection criteria are applied on the catalogue, and two subsets are defined: one for training the extinction model, and one for evaluating it (Section 5.1).
(iii) Linear interpolation is used for each source to find the best theoretical spectra in the 4D grid of the stellar parameter space (Section 5.2).
(iv) The atmosphere chromatic transmission is simulated for each source at each observation epoch using LIBRADTRAN (Section 5.3).
(v) Theoretical top-of-atmosphere (TOA) and through-the-atmosphere magnitudes are calculated from the synthetic SEDs, the instrumental transmission and the synthetic atmospheric transmission. The atmospheric cloud-free extinction is calculated from the difference between the TOA and under-the-atmosphere theoretical magnitudes. These corrections are added to the measured magnitudes to get the TOA observed magnitudes (Section 5.4).
(vi) The sequence with the brightest observations, assumed to be photometric, is identified as the reference sequence. The reference TOA magnitudes of the sources are estimated by calculating the average of TOA observed magnitudes in this sequence.
(vii) $\mathrm{zp}^{\rm obs}$, the instrumental zero-point acting as a global offset, is obtained by a linear fit of the reference TOA magnitudes to the *Gaia* converted SDSS-$r$ magnitudes from equation (2) (see Fig. 5). Sources with residual above $3 \times \sigma_{\rm Gaia}$ are removed from the data set.
(viii) Differences between the TOA observed magnitudes and reference TOA observed magnitudes yield the estimated grey extinction for each source (Section 5.5).
(ix) Grey extinction correction models are adjusted on the grey extinction values from the training subsets (Section 5.6).

### 5.1 Selection criteria and catalogue of objects

Once forced photometry had been performed we applied the following selection criteria:

(i) To ensure the accuracy of the photometric catalogue, bright and unsaturated stars were selected with $8.5 \leq r \leq 13.5$ out-of-atmosphere magnitude, calculated from *Gaia* DR3 photometry via equation (2). This corresponds to stars with an SNR ranging from 3 to 300 depending on atmospheric conditions and sky background levels.
(ii) Airmass: $1 \leq X \leq 2$
(iii) Effective temperature: $3500\,\mathrm{K} \leq T_{\rm eff} \leq 8000\,\mathrm{K}$
(iv) Surface gravity: $0.0 \leq \log g \leq 5.5$
(v) Metallicity: $-1.5 \leq [\mathrm{Fe/H}] \leq 0.5$
(vi) Nearest neighbours angular separation is at least 15 arcsec
(vii) Sources identified as variable in the *Gaia* catalogue are excluded (`VarFlag = False`)
(viii) Images with an aperture correction value $\Delta_{i,k} < -0.3$ mag are excluded.

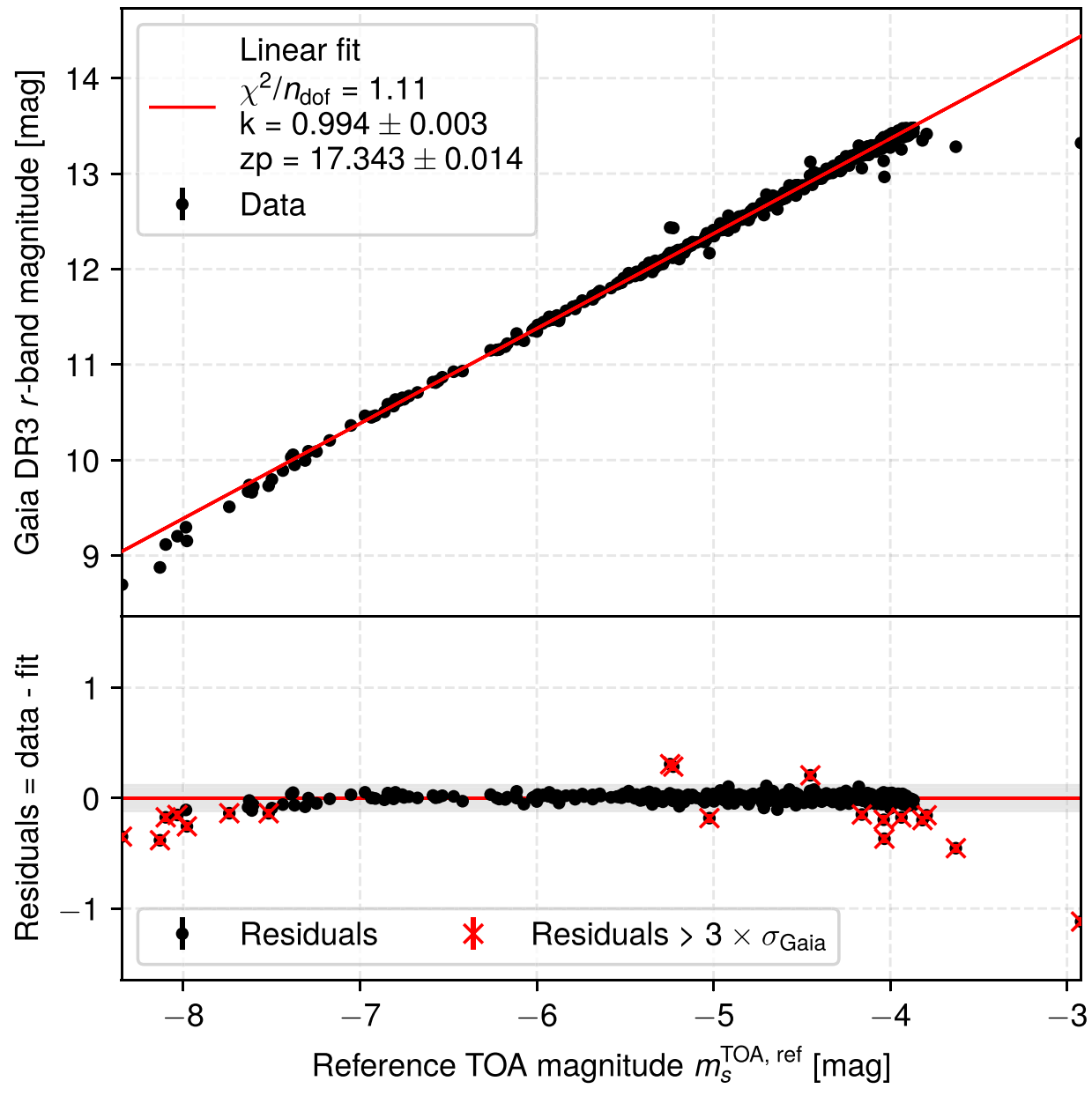

**Figure 5.** Linear relation between 'reference' top-of-atmosphere magnitudes of all the stars in our data set and the transformed *Gaia* DR3 $r$-band magnitudes computed with equation (2). The linear curve is fitted using a robust least-squares method with a smooth approximation to absolute value loss function (`soft_l1`) to prevent contamination from a small amount of outliers. The grey shaded area on the lower panel is delimited by $\pm\sigma_{\rm Gaia} = 0.03776$ mag, which corresponds to the *Gaia* DR3 photometric system conversion function uncertainty. Stars identified by red crosses are excluded from the catalogues.

(ix) Sources whose centroid, calculated by the centre of mass method, is more than 5 pixels away from the known position based on the WCS reference transformation are excluded.

All stars meeting these selection criteria for each field were randomly divided into two subsets following a train-test split scheme. The first subset, called the 'training' set, contains 70 per cent of the sources used to adjust the parameters of the grey extinction models. These are the stellar calibrators. The second subset, called the 'test' set, contains the remaining 30 per cent of the sources and is used to evaluate the accuracy of the model.

### 5.2 Construction of SED lookup tables

We use synthetic spectrum models from the AMBRE collection (P. de Laverny et al. 2012), available in open access[12] on the POLLUX data base (A. Palacios et al. 2010). AMBRE corresponds to a grid of synthetic spectra across the entire optical domain for cool stars ($3500 \leq T_{\rm eff} \leq 8000$ K) of all luminosities (from dwarfs to supergiants), with metallicities ($-5.0 \leq [\mathrm{Fe/H}] \leq 1$ dex) and varying surface gravities ($-0.5 \leq \log g \leq 5.5$ dex), generated using the MARCS atmospheric model grids (B. Gustafsson et al. 2008). For each source $s$ identified by its *Gaia* ID, we associate an interstellar dust extinguished synthetic spectrum outside the atmosphere, at the star's surface, through linear interpolation of the 4D grid ($\lambda$, $T_{\rm eff}$, $\log g$, [Fe/H]).

[12]The POLLUX Data base: https://pollux.oreme.org/

To account for interstellar dust extinction along the line-of-sight, we apply reddening corrections to the SEDs of each source. This is done by combining the positional data from *Gaia* DR3 with the Schlegel, Finkbeiner & Davis (SFD) dust map (E. F. Schlafly & D. P. Finkbeiner 2011) and the E. L. Fitzpatrick et al. (2019) reddening law. We implement these corrections using the DUST_EXTINCTION[13] package (K. D. Gordon 2024) in conjunction with DUSTMAPS[14] (G. Green 2018). The colour excess $E(B-V)$ is extracted from the SFD map based on the source positions, and we adopt a standard total-to-selective extinction ratio of $R_V = A_V/E(B-V) = 3.1$ to model the extinction of the synthetic spectra.

### 5.3 Simulation of the atmosphere transmission

Stellar light is affected by atmospheric extinction, that includes both chromatic and achromatic effects (C. W. Stubbs & J. L. Tonry 2006; C. W. Stubbs et al. 2007; D. L. Burke et al. 2010, 2013, 2017):

(i) Line absorption, primarily caused by three molecules: oxygen ($O_2$), water vapour ($H_2O$) – represented by its precipitable amount (Precipitable Water Vapour, hereafter PWV) – and ozone ($O_3$) (A. Bucholtz 1995).
(ii) Continuous Rayleigh scattering by molecules such as $O_2$, $O_3$, $H_2O$, $N_2$, and trace elements.
(iii) Mie scattering (D. Jackèl & B. Walter 1997) by aerosols (e.g. dust, sand, sea salt, soot), with physical dimensions comparable to UV or visible light wavelengths.
(iv) Grey extinction caused by large ice crystals and water droplets in clouds, independent of wavelength.

Details about the non-grey components of the atmosphere transmission are given in Appendix C.

For an observation labelled $i$, performed at an epoch $t_i$, we write the set of parameters required to model the atmosphere as

$$\Theta_i^{\mathrm{atm}} = \{t, \mathrm{alt}, \mathrm{az}, X, \mathrm{P}^{\mathrm{atm}}, \mathrm{PWV}, \mathrm{O}_3, \tau_0^{\mathrm{aer}}, \beta^{\mathrm{aer}}, \mathcal{P}^{\mathrm{season}}\}_i \quad . \tag{7}$$

This set includes the telescope pointing altitude alt and azimuth az, the airmass $X$, the atmospheric pressure at the altitude of the observation site $\mathrm{P}^{\mathrm{atm}}$, the precipitable water vapour content PWV, the total vertical ozone column $O_3$, the aerosol optical depth $\tau_0^{\mathrm{aer}}$ at $\lambda_0 = 500$ nm and Angström exponent $\beta^{\mathrm{aer}}$, and the vertical profile of temperature, relative humidity and other gases $\mathcal{P}^{\mathrm{season}}$.

The spectral terrestrial atmospheric transmission at a wavelength $\lambda$ taken in the near-UV to near-IR (300–1100 nm) range can then be written as the product of a chromatic and a grey term:

$$\mathcal{T}^{\mathrm{atm}}(\lambda, \Theta_i^{\mathrm{atm}}) = \mathcal{T}^{\mathrm{chromatic}}(\lambda, \Theta_i^{\mathrm{atm}}) \times \mathcal{T}^{\mathrm{gray}}(t_i, \mathrm{alt}_i, \mathrm{az}_i) \quad . \tag{8}$$

Apart from the aerosol component, chromatic atmospheric transmission is highly predictable and varies slowly over time and short spatial scales (C. W. Stubbs et al. 2007; A. Boucaud 2013). Thus, we consider in this work that the chromatic components of the atmospheric transmission contained in the right-hand side of equation (8) have no line of sight spatial dependence and thus does not depend on azimuth. Therefore, only the airmass is used to compute these quantities. Conversely, the grey term being dependent on the cloud cover and structure is highly dependent on azimuth and altitude.

[13]Version 1.7: https://dust-extinction.readthedocs.io/en/latest/
[14]https://dustmaps.readthedocs.io/en/latest/

The chromatic atmospheric transmission $\mathcal{T}^{\mathrm{chromatic}}(\lambda, \Theta_i^{\mathrm{atm}})$ is linearly interpolated for each star per observation epoch over a high-resolution 4D grid – generated with LIBRADTRAN – in ($X$, $\mathrm{P}^{\mathrm{atm}}$, PWV, $O_3$), airmass, air pressure, water vapour, and ozone total column with the following parameter space: $1 \leq X \leq 2$ with $\Delta X = 0.1$, $900 \leq \mathrm{P}^{\mathrm{atm}} \leq 1100$ with $\Delta \mathrm{P}^{\mathrm{atm}} = 5$ hPa, $0 \leq \mathrm{PWV} \leq 50$ with $\Delta\mathrm{PWV} = 1$ mm and $200 \leq \mathrm{O}_3 \leq 400$ with $\Delta \mathrm{O}_3 = 25$ DU. In Fig. C1, we depict such a chromatic atmosphere transmission curve with individual components for OHP average conditions given in the caption.

### 5.4 Predicted atmospheric chromatic correction to broad-band photometry

In order to derive the grey extinction we first correct the instrumental magnitude $m_{s,i}^{\mathrm{inst}}$, defined in equation (3), from atmospheric chromatic extinction $\mathcal{C}_{s,i}$, to yield the TOA instrumental magnitude:

$$m_{s,i}^{\mathrm{TOA,\,inst}} = m_{s,i}^{\mathrm{inst}} - \mathcal{C}_{s,i} \,, \tag{9}$$

where

$$\mathcal{C}_{s,i} = 2.5\,\log_{10}\left(\int_0^\infty \mathcal{S}_s(\lambda, T_{\mathrm{eff}}, \log g, \mathrm{Fe/H}) \times \mathcal{T}_b^{\mathrm{inst}}(\lambda) \times \left[1 - \mathcal{T}^{\mathrm{chromatic}}(\lambda, \Theta_i)\right] \times \frac{\lambda}{hc} \times \mathrm{d}\lambda\right) , \tag{10}$$

with $\mathcal{S}_s(\lambda, T_{\mathrm{eff}}, \log g, \mathrm{Fe/H})$ the object SED (attenuated by interstellar dust) and $\Theta_i$ defined in equation (7). $\mathcal{T}_b^{\mathrm{inst}}(\lambda)$ is the instrumental transmission calculated based on measurements provided by the manufacturers or third parties.[15] It is defined as (see also Fig. 3)

$$\mathcal{T}^{\mathrm{inst}}(\lambda) = \mathcal{T}^{\mathrm{opt}}(\lambda) \times \mathcal{T}^{\mathrm{filter}}(\lambda) \times QE(\lambda) \,, \tag{11}$$

where $\mathcal{T}^{\mathrm{opt}}(\lambda)$ is the total optical transmission of the telescope (corresponding in this case to the product of the apochromatic doublet lenses and the flat-field corrector[16] transmissions), $\mathcal{T}^{\mathrm{filter}}(\lambda)$ is the $r$ filter bandpass and $QE(\lambda)$ is the absolute quantum efficiency of the detector. These instrumental transmission curves all considered independent of the geometric position in the focal plane, and we assume that the observation campaign duration is short enough that we can neglect time variations of the instrumental transmission. The impact of stellar parameters and instrumental transmission errors on $\mathcal{C}_{s,i}$ is assessed and discussed in Appendix D.

### 5.5 Grey extinction determination

The grey extinction $\Delta m_{s,i}$ for each source $s$ in an image $i$ is calculated as the difference between the instrumental magnitude, corrected from all but grey extinction to the top-of-atmosphere, $m_{s,i}^{\mathrm{TOA,\,inst}}$ (equation 9) and its reference magnitude $m_s^{\mathrm{TOA,\,ref}}$,

$$\Delta m_{s,i} \equiv m_{s,i}^{\mathrm{TOA,\,inst}} - m_s^{\mathrm{TOA,\,ref}}. \tag{12}$$

[15]The refractor's optical transmission is almost grey in the $r$ filter wavelength range, according to the refractive index data base (M. Polyanskiy 2024) and to the FPL-53 and Lanthanum lenses used in the apochromatic telescope: https://refractiveindex.info/?shelf=glass&book=OHARA-FPL&page=FPL53
[16]Following S. Garrappa et al. (2025), due to the lack of measurements for the flat-field corrector, we assume that its transmission exhibits a similar wavelength dependence to that of the optical tube assembly lenses.

This reference magnitude is defined as the average (with a 2.5-$\sigma$ clipping) of all observations obtained under the best photometric conditions and belonging to a unique observation night per target field, which we will refer to as the reference catalogue. This observation night sequence is selected beforehand as the one exhibiting the lowest dispersion and lowest (brightest) average TOA instrumental magnitudes (equation 9), for each target field (see Tables A2 and A3).

As explained by D. L. Burke et al. (2017), the grey extinction term $\Delta m_{s,i}$ is a combination of multiple effects, including: instrumental effects (dome occultation, shutter timing error), residuals in assignments of ADU counts to celestial sources, modelizations errors of $\mathcal{C}_{s,i}$ (due to LIBRADTRAN simulations and input parameters errors) and grey cloud extinction. Some of these effects may depend on the filter band, and therefore show some chromaticity as well as being small compared to the grey extinction caused by clouds. At this stage, we cannot verify the chromaticity with only one filter bandpass. Therefore, for the remaining of the analysis, we assume $\Delta m_{s,i}$ to be primarily grey and caused by clouds and chromatic effects other than those evaluated in the previous section to be negligible.

### 5.6 Radiometric model

We outline here the method for fitting the grey extinction model on the training subset by comparing $\Delta m_{s,i}$ (equation 12) and the radiance excess $\Delta L_{s,i}$ (equation 6). After adding a constant parameter $C$ to the relation linking visible extinction and IR radiance given in equation (1) and inverting it, yields,

$$\hat{\Delta} m_{s,i} = -\frac{1}{B_i} \times \log\left(\frac{A_i + C_i - \Delta L_{s,i}}{A_i}\right), \quad (13)$$

where $A_i$ and $B_i$ are positively constrained fitted parameters. Whenever the IR radiance excess is too low (i.e. the extinction remains small), only a linear trend appears. In this case, is not possible to properly constrain all parameters of the physical model as they are highly correlated and we fallback to a simpler model:

$$\hat{\Delta} m_{s,i} = A_i \times \Delta L_{s,i} + B_i. \quad (14)$$

These parameters are adjusted for each image $i$ through robust least-squares minimization of residuals $R_{s,i} = \Delta m_{s,i} - \hat{\Delta} m_{s,i}$ using the objective function from J. Tellinghuisen (2020) considering uncertainties in both $\Delta m_{s,i}$ (y-axis data) and $\Delta L_{s,i}$ (x-axis data),

$$\chi_i^2 = \sum_s \frac{R_{s,i}^2}{\sigma(\Delta m_{s,i})^2 + \left(\frac{\mathrm{d}\hat{\Delta} m_{s,i}}{\mathrm{d}\Delta L_{s,i}}\right)^2 \times \sigma(\Delta L_{s,i})^2}. \quad (15)$$

Measurement uncertainties on the measured radiance $\sigma(L_{s,i})$ are taken as the standard deviation of the stack measurement per pixel and uncertainties on the initial estimate of grey extinction are taken as the photometric uncertainty from equation (4) such as $\sigma(\Delta m_{s,i}) \equiv \sigma^2(m^*_{s,i})$.

We use LMFIT[17] (M. Newville et al. 2025) with `soft_l1` loss function because the data is sometimes contaminated with outliers that cannot be removed by other means. The Trust Region Reflective algorithm was chosen for minimization due to its robustness for problems with bound constraints.

[17] Version 1.3.4: https://lmfit.github.io/lmfit-py/

### 5.7 Corrected magnitudes, errors, and uncertainties

Once the best-fitting model is obtained, the TOA instrumental magnitudes of all calibration and test objects are corrected, to yield the corrected observations $m_{s,i}^{\mathrm{TOA,\,corr}}$,

$$m_{s,i}^{\mathrm{TOA,\,corr}} = m_{s,i}^{\mathrm{TOA,\,inst}} - \hat{\Delta} m_{s,i}, \quad (16)$$

where $\hat{\Delta} m_{s,i}$ is the grey extinction model prediction.

The calibration errors $\epsilon_{s,i}$ for each source $s$ in each image $i$ can then be computed as,

$$\epsilon_{s,i} = m_{s,i}^{\mathrm{TOA,\,corr}} - m_{s}^{\mathrm{TOA,\,ref}}. \quad (17)$$

This is the difference between the observed magnitude extrapolated at the top-of-atmosphere and corrected for grey extinction (equation 16), with the 'reference' magnitude.

## 6 RESULTS

For the majority of exposures, there is either no grey extinction or the data is too sparse to reliably fit the non-linear deviation that appears at high values of grey extinction $\Delta m_{s,i}$. Instead, only a linear relationship between grey extinction and radiance excess is observable for nearly half of them. As a result, fitting the full physical model often yields unstable behaviour, with parameters exhibiting unrealistically large covariances. To address this issue for images acquired under low but non-zero levels of grey extinction, we instead fit a simple linear model (equation 14) – with a positively bounded slope – whenever the relative uncertainty of either parameter $A$ or $B$ in the physical model exceeds 50 per cent. If the fitted slope hits the lower bound of zero (or the uncertainty straddles around zero), no correction is applied to the observed magnitudes, as this indicates an absence of grey extinction in the image. Overall, approximately 5 per cent of the images are corrected using the physical model described in equations (1) and (13) (group n°1), ~48 per cent are corrected by the linear model (group n°2) and the remaining ~47 per cent do not contain sufficient grey extinction to be corrected for (group n°3). To illustrate the behaviour of these three groups, we selected three exposures representing the three group types described above. Diagnostic plots for these cases are shown in Figs 6, 7, and 8 and results are summarized in Table 1. Top-left panels show the map of radiance excesses with black dots indicating the stars positions and the white rectangle the footprint of the CMOS image. Top-right panels show the measured grey extinction values for each star of the training (black errorbars) and test (grey errorbars) subsets plotted against the radiance excesses $\Delta L_{s,i}$ with the best-fitting curve in red. The lower left and right panels of each figure shows respectively the estimated and inferred grey extinction values from the fit for test sources as a function of CMOS detector coordinates. The numbers indicated in the upper-left corner are the calculated dispersion of the photometric errors (equations 12 and 17).

Results for exposure n°2548 (Fig. 6) indicate that the stars are observed through a large average grey extinction of 1.54 mag on the night of July 12. It can be noticed that the grey extinction follows a fairly linear variation with radiance excess from ~0.4 to ~1.5 mag (Fig. 6b), before reaching the non-linearity regime from 1.5 to 3.0 mag. Looking at the left column plots, we see a smooth left to right gradient of radiance excess (grey extinction) on the image. The dispersion of the error on individual TOA magnitude measurements is reduced from 0.732 mag before correction to 0.152 mag after fitting and applying the model. The use of `soft_l1` loss helps mitigating the influence of a single outlier that cannot be removed by other means, located at approximately 3.9 W m$^{-2}$ sr and 2.5 mag, preventing the fit from being distorted.

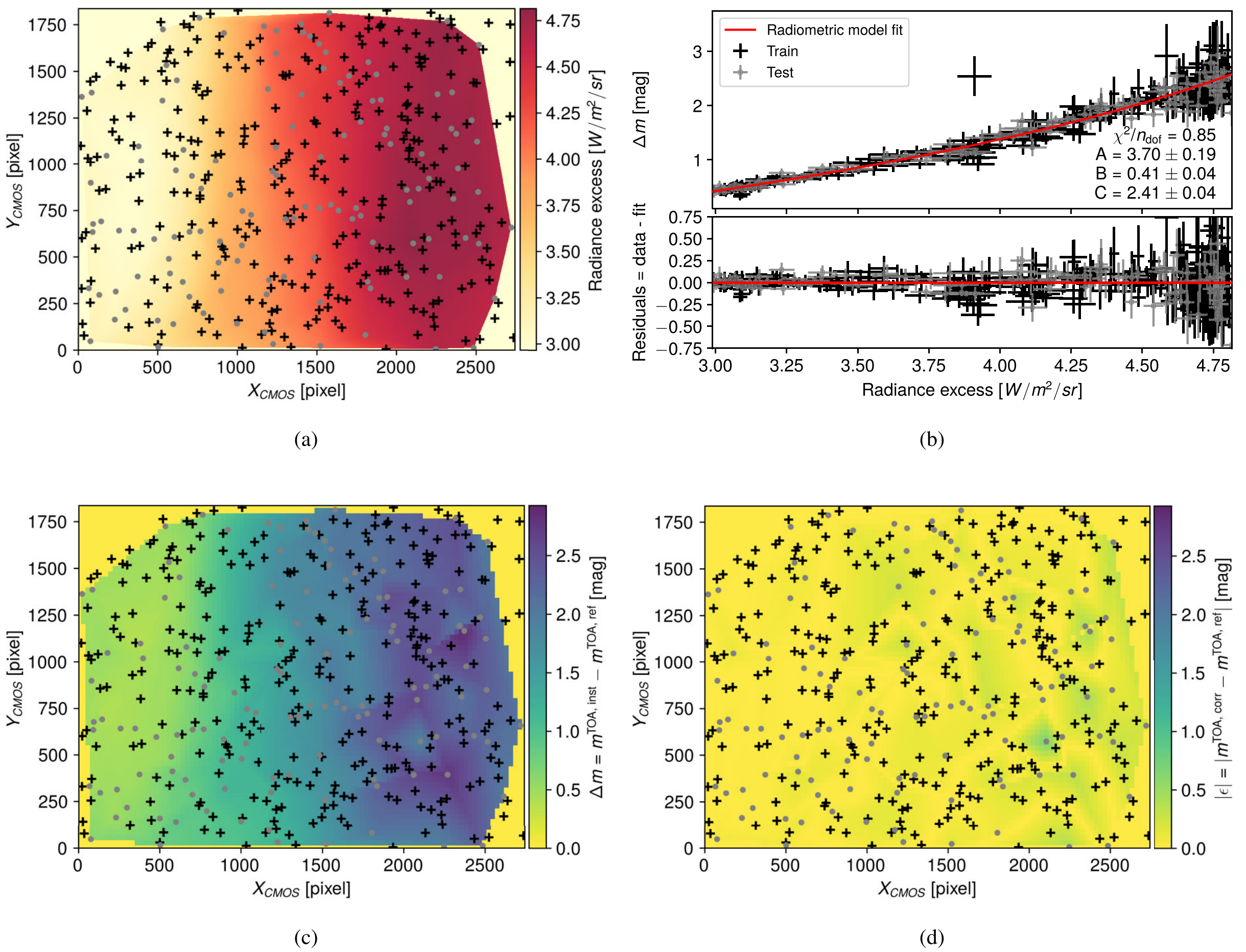

**Figure 6.** Results for image 2548 of the target field BD+28 4211 observed on the night of July 12. Panel (a): Map of radiance excess projected onto the CMOS pixels coordinates grid. The radiance excess is shown for train ($N_{\mathrm{train}} = 294$; black crosses) and test stars ($N_{\mathrm{test}} = 125$; grey circles) and linearly interpolated on a grid. Panel (b): Best-fitting curve (solid red) of the grey extinction physical model versus radiance excess. The lower panel displays the residuals. Panel (c) Grey extinction initial estimates (equation 12) as a function of CMOS image coordinates. The colourmap corresponds to the linear interpolation of test stars values on a grid. Panel (d): Idem for the absolute value of residual errors (equation 17) on corrected magnitudes. The colourbar range is kept identical to ease comparison.

For exposure n°2538 (Fig. 7), the range of grey extinction is between 0.7 and 1.7 mag. The relation between radiance excess and grey extinction is well fitted by the linear model with well constrained parameters with small uncertainties (see Fig. 7b). Applying the model on test stars leads to a reduction of the dispersion on corrected magnitude errors by a factor 3 from 0.244 to 0.080 mag.

Exposure n°2329 (Fig. 8) seems to be dominated by photometric and radiometric noise. The application of a linear correction is pointless (see Fig. 8b) as photometric measurements are scattered across null extinction and the radiance excess amplitude of 0.25 W m$^{-2}$ sr is below the 3-$\sigma$ noise and detection limit of our IR camera (around 0.1 W m$^{-2}$ sr at 1-$\sigma$; K. Sommer et al. 2024). The fitted slope parameter of 0.01 ± 0.06 is consistent with zero.

### 6.1 Correction accuracy

To quantify the accuracy of the correction models, we estimate the propagated error on corrected magnitudes using the bootstrap method. As we know that the parameters asymptotically follow a multivariate normal distribution, we simulate a large number ($N_{\mathrm{sim}} = 1000$) of new experiments with varied parameter values and compute the standard deviation of these simulated realizations to estimate the models accuracies. The distributions of the propagated uncertainties of the linear (green curve) and physical models (red curve) corrections are shown in Fig. 9 and are clipped to a minimum and maximum boundary at 1 mmag and 0.15 mag respectively. Values outside of this range are clipped to the given interval edges. They both have a Gamma distribution shape with values ranging from 1 to 150 mmag for the high-end tail. The mean values are 5 and 29 mmag for the linear and the physical model respectively. This can be explained by the fact that the physical model is used to correct for large extinctions. When examining the percentage ratio $\sigma^{\mathrm{model}}/\sigma^{\mathrm{phot}}$ between models propagated error and the photometric noise for each source (equation 4), both curves peak at 5 per cent and the mean values are 10 per cent and 21 per cent for the linear and physical models respectively. This emphasizes the dominance of photon noise in the total uncertainty budget and the accuracy of the linear model being superior to the physical model.

In Fig. 10, we show the standard deviation of differences between individual TOA magnitudes and reference magnitudes for observations of all test sources belonging to the groups 1 and 2 before (blue points; equation 12) and after applying the grey extinction correction (orange triangles; equation 17). Uncorrected observations classified during good atmospheric conditions and thus estimated free from measurable grey extinction are shown as grey crosses for comparison. Two model curves representing the photometric uncertainty noise model (red) and the lower limit of the correction method (green) are superimposed for comparison. The photometric noise uncertainty model is constructed by fitting a spline to the average of photometric uncertainties measured on the reference nights as a function of the stars respective $r$-band magnitudes computed with equation (2). The scatter displays a convex shape for corrected observations that ranges from $\sim$0.10 mag for the brightest sources with $r \leq 9.0$ to $\sim$0.07 for intermediate brightness $10 \leq r \leq 12$ and back up to $\sim$0.10 for low brightness $r \geq 12.5$. The standard deviation for uncorrected observations follow a similar trend. The increase of standard deviation for low magnitudes stars (i.e. high brightness) with $8.5 \leq r \leq 10$ is due to saturation. The improvement in errors dispersion after correction of group 1 and 2 data is clear, with a $\sim$2–5 times reduction. Group n$^\circ$3 data has standard deviations higher than corrected data from group n$^\circ$1 and 2. Again, some images from group n$^\circ$3 are affected by weak grey extinction that we cannot measure because it corresponds to a signal below the noise detection threshold of either the radiometric instrument and/or photometric observations. We attempt to correct these observations using a constant value per exposure, defined as the weighted average of grey extinction estimates of training sources. It does improve the correction efficiency and standard deviations follow the trend of the photometric uncertainty model.

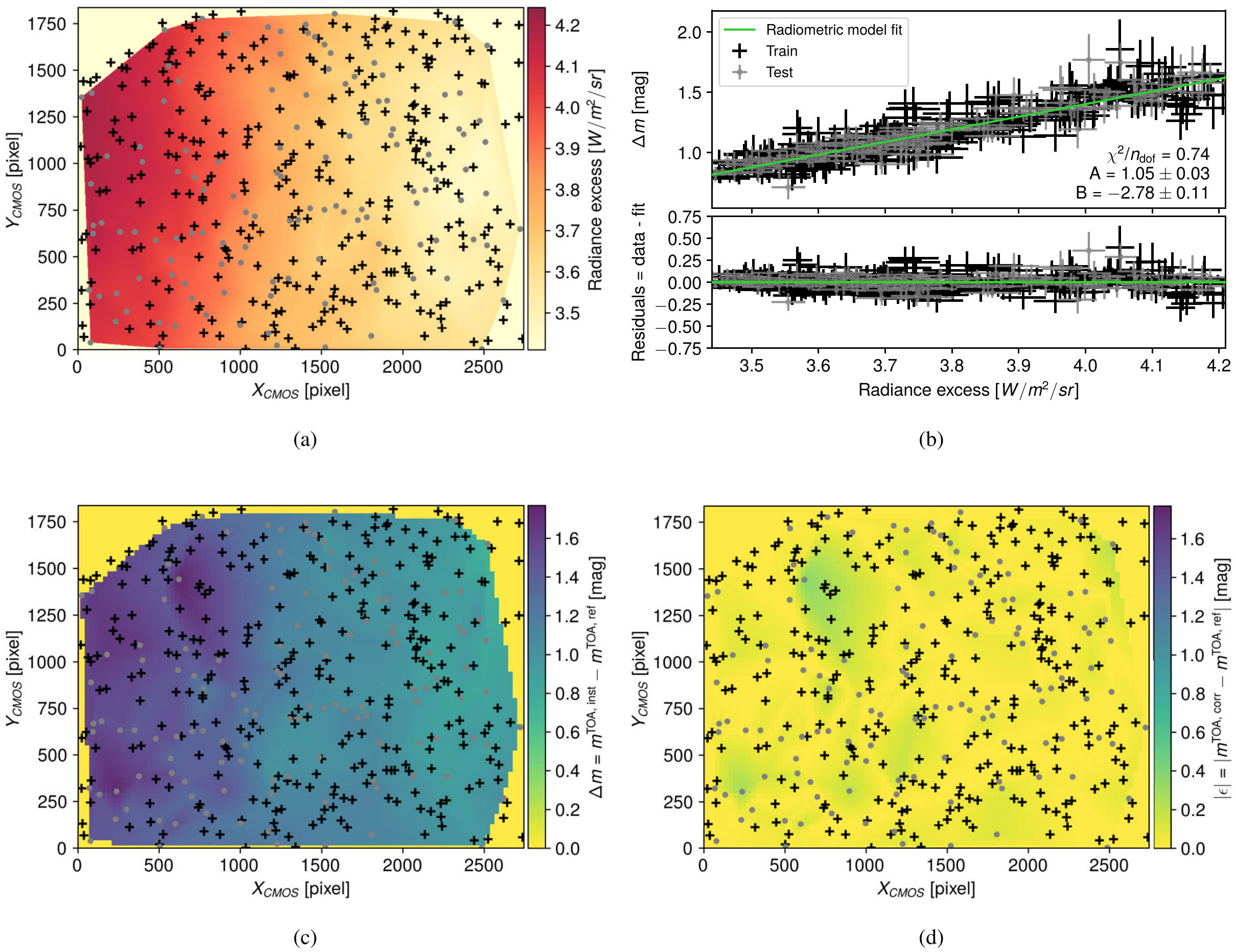

**Figure 7.** Same as Fig. 6 for image 2538 of the BD+28 4211 target field observed on the night of July 12. This time, the shown radiometric correction model is the linear fit (green curve) as the absence of non-linearity in the data prevents the physical model to be accurately fitted.

### 6.2 Temporal stability

To demonstrate the temporal stability of the radiometric correction model, we randomly selected two non-variable test sources of $r$ magnitudes equal to 11.835 and 10.925 respectively for each field across two observation sequences of over 1200 observations in total, under poor to average atmospheric conditions. In Fig. 11, we show the light curves before correction (blue), along with the corrected data (orange). Visual inspection of the light curves, and comparison of the displayed standard deviation show that the correction model systematically reduces the scatter. It provides good corrections even in the case of large (up to 3 mag) extinction.

Even in cases of subtle grey extinction, which introduces a slight slope in the observed magnitudes, for example, the second part of the night of July 24 – the model successfully corrects the data. For the night of August 4, the scatter in the extinction-corrected magnitudes approaches the intrinsic photon noise level of 0.018 mag. While the correction quality decreases with increasing extinction, it remains robust, yielding an RMS scatter of approximately 0.134 mag even under severe conditions, such as on the night of July 20, when extinction reached up to 4 magnitudes for a significant portion of the sequence.

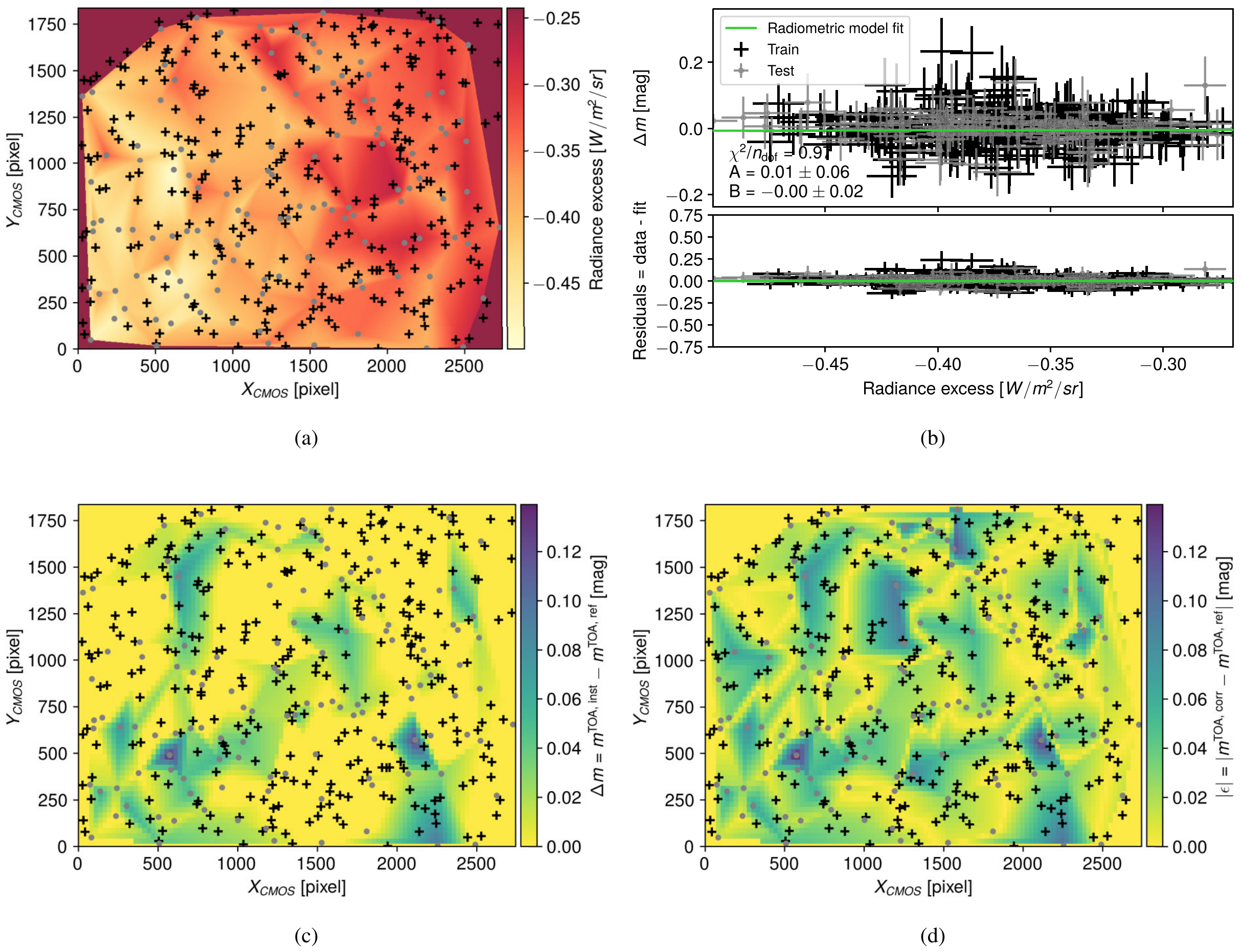

**Figure 8.** Same as Fig. 7 for image 2329 of the BD+28 4211 target field observed on the night of July 12. For this case, no grey extinction does appear and the fitted linear model slope value confirms it. The radiance gradient in (a) follows the airmass gradient.

**Table 1.** Summary of the results presented in Figs 6, 7, and 8. The following quantities are listed: (1) the exposure number index; (2) the model used to correct photometric data; (3), (4), and (5) the best-fitting model parameters; (6) and (7) the standard deviation of the photometric errors before (equation 12) and after correction (equation 17).

| Exp. index | Model | A | B | C | $\sigma$ before correction | $\sigma$ after correction |
|---|---|---|---|---|---|---|
| 2548 | Physical | $3.70 \pm 0.19$ | $0.41 \pm 0.04$ | $2.41 \pm 0.04$ | 0.732 | 0.152 |
| 2538 | Linear | $1.05 \pm 0.03$ | $-2.78 \pm 0.11$ | N.A | 0.244 | 0.080 |
| 2329 | Linear | $0.01 \pm 0.06$ | $0.00 \pm 0.02$ | N.A | 0.042 | 0.042 |

## 7 CONCLUSIONS

Using a calibrated infrared thermal camera performing concurrent radiometric measurements to photometric observations, we have successfully demonstrated the feasibility to correct ground-based astronomical observations for grey extinction. The method consist of deriving a grey extinction model for each exposure expressed as a function of radiance excess and optical depth. After applying it to an observation campaign, it enabled corrections of temporal variations of the extinction in not perfectly photometric sequences (e.g. affected by as much as 3.0 magnitudes of $r$ extinction) at the level of 0.025 mag with a resolution of 1 arcmin, which is encouraging for its use case in StarDICE. When highly variable conditions are sparse over the night, the corrected photometry shows a scatter slightly higher than the photon noise (Fig. 11). A key innovation is our process to separate atmosphere grey and chromatic extinction by combining a forward model with auxiliary atmospheric data (PWV, ozone, aerosols, etc.), Earth's atmosphere radiative transfer simulations and synthetic photometry calculations using *Gaia* DR3 data, which led to accurate modelling of atmospheric effects and sky down-welling radiance, allowing precise inference of grey extinction from thermal images. This is, to our knowledge, the first quantitative application of thermal imaging for this purpose, previously proposed by D. L. Burke et al. (2013) and F. Hazenberg (2019).

Despite spanning a large range of atmospheric conditions, our observation campaign certainly does not include all possible types or thickness of clouds. It was carried at a site with higher water vapour columns than at large observatories as, for example, in Chile. Nevertheless, the implemented setup and collected data do allow a number of important general issues to be addressed, and some conclusions on the methods to be drawn. We can also suggest potential future improvements and use-case scenarios.

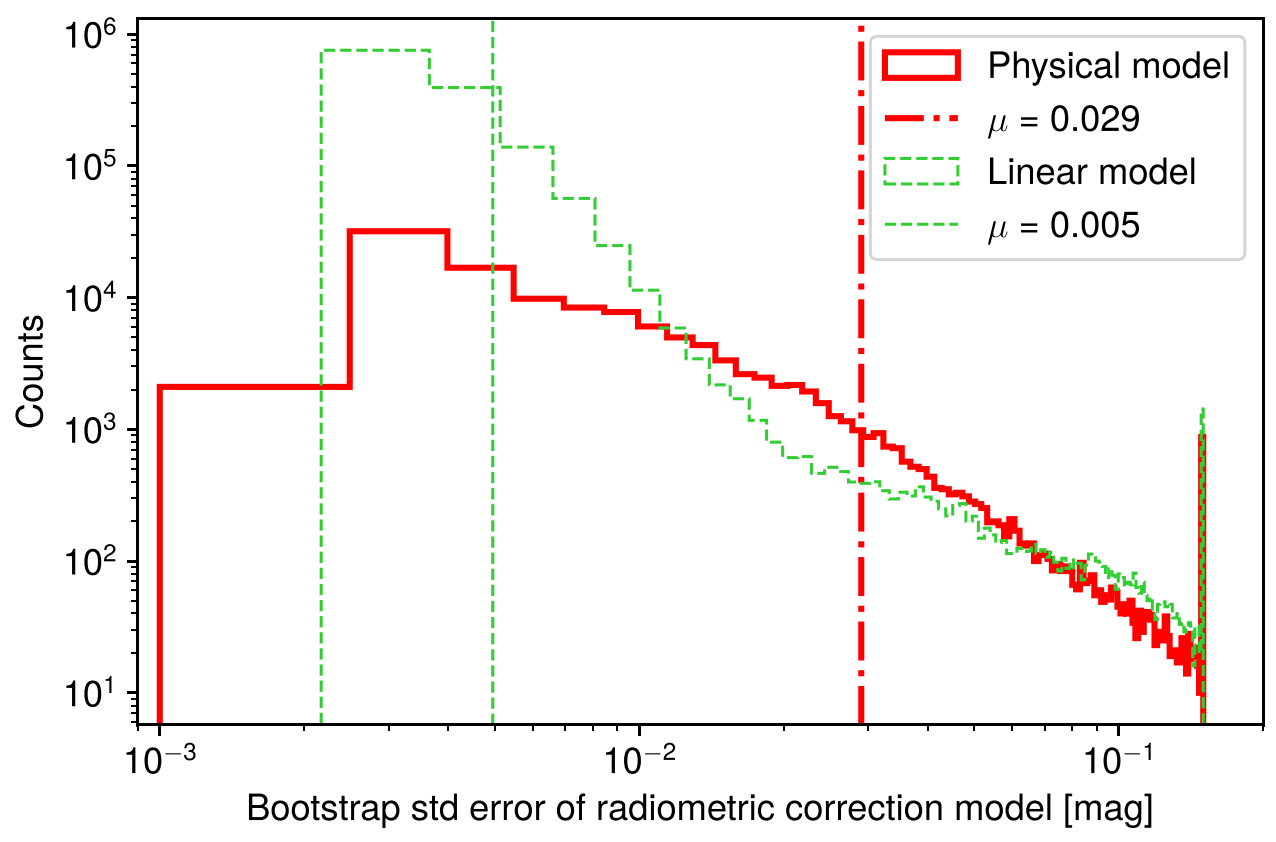

**Figure 9.** Clipped distributions of the bootstrap standard error of the correction model predictions for all sources and exposures belonging to groups n°1 and n°2. Overflow and underflow counts are shown in the edge bins of the plot. Errors larger than 0.15 mag represent less than one per cent of the data. The coloured vertical lines mark the unclipped averages of the distributions.

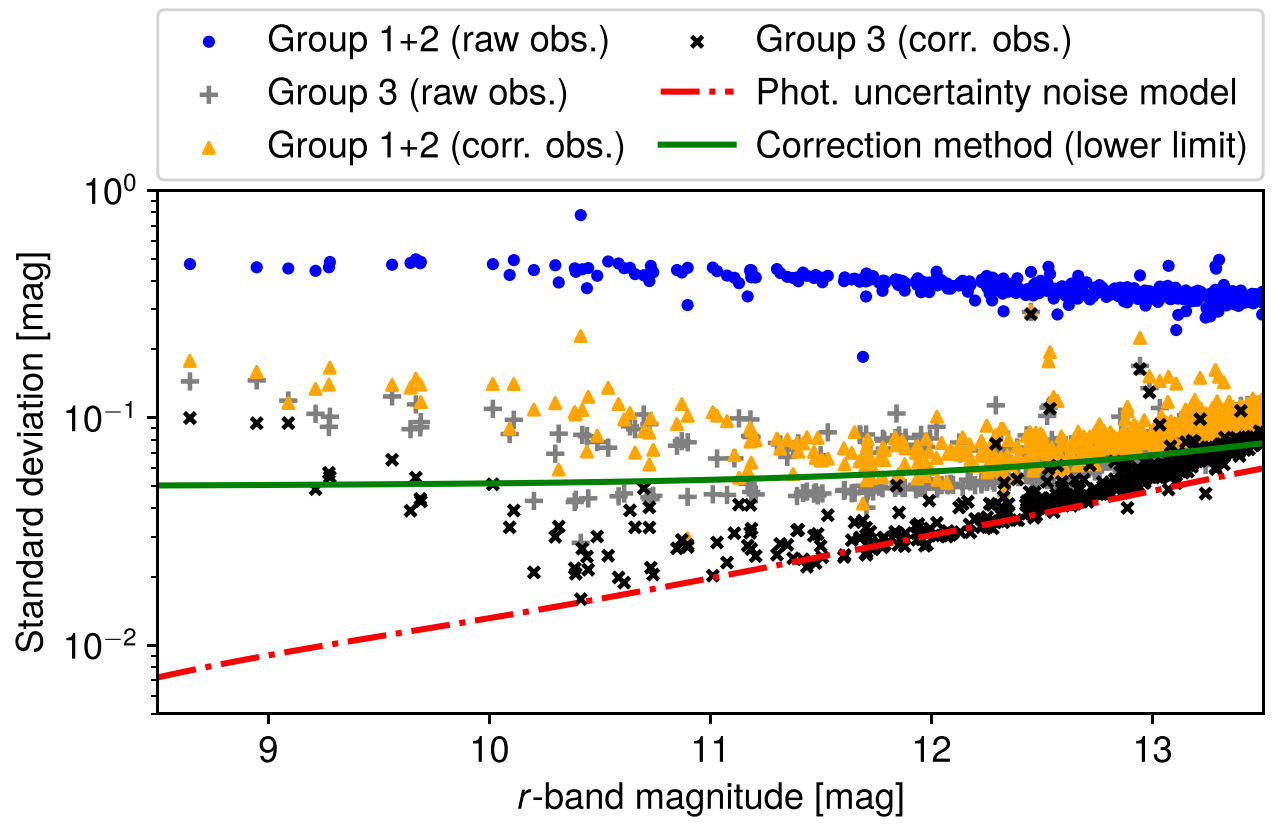

**Figure 10.** Standard deviation of errors (equation 17) as a function of the $r$ magnitude for all test sources belonging to the target field BD+28 4211. The blue points depict the raw observations belonging to the combination of groups n°1 and n°2, impacted by grey extinction, whereas grey crosses belong to the group n°3. Orange triangles represent the standard deviation of corrected data errors of groups n°1 and n°2. Black crosses represent those from group n°3 corrected by a constant for each image. The red dashdotted curve is a spline model of photometric uncertainties as a function of magnitude determined from the reference photometric sequence. The green solid curve is a combination of the latter and a constant limit fixed at 50 mmag.

Our analysis uses only stacked IR frames within a single optical exposure, and does not take advantage of the high-speed imaging capability of the IR camera as thermal instabilities limit the use of temporal information. Advanced processing techniques, for example, relying on structure detection, could bring useful information on the grey extinction evolution within a single optical exposure. This is probably necessary if the technique is used for longer exposure optical images, with various types of clouds moving over large portions of the field.

Paths to improve the instrumental setup are clear after identification of a number of drawbacks and issues. For 20 s exposure, the magnitude reachable with the current optical tube assembly (collection area of 40 cm$^2$) is approximately $r = 13.5$. Therefore, the usable number of stars per image is less than ∼450 on average. This limits the precision of our grey extinction correction model. It also prevents us from directly comparing our results with the photometric model correction method of D. L. Burke et al. (2013) and S. Garrappa et al. (2025). Indeed, when attempting to fragment our images into smaller patches to fit grey extinction residuals using 2D polynomial functions, we found that our data set did not provide a sufficient number of calibration stars per patch to enable a reliable unbiased comparison with the one dimensional radiometric model. In particular, the limited star density hindered the robust fitting of six-parameter polynomials. However, this limitation further emphasizes the advantage of the proposed radiometric model for such case scenario, which leverages the full image content to correct for grey extinction. It does not suppose any structure, scale, or smoothness of the extinction contrary to the photometric method mentioned above, which requires to assume a smooth variation of extinction across an image patch.

The sampling resolution of the infrared thermal camera (58.7 arcsec pixel$^{-1}$) implies that a single IR pixel may contain more than one optical source. It is not possible to probe the grey extinction structure function at a scale below about 2 arcmin with our camera (∼6 m at 10 km height). The athermalized lens could be replaced by a 100 mm version, almost doubling its resolving power. It would require another complete in-laboratory radiometric calibration, as detailed in K. Sommer et al. (2024), and would reduce the field with the same factor.

It was found that the close thermal environment of the IR camera is a major source of noise. Prior to collecting data for this study, the instrument was removed from a dome as relative rotations of the mount and dome induced thermal fluctuations that we did not succeed to account for. It was then installed in a roll-off-roof building to ensure a close environment as stable as possible. Carefully designed aluminium baffling could further reduce environmental noise.

Finally, only a single $r$ photometric filter was used in this study. We are thus not able to fully conclude on the actual achromaticity of the grey extinction correction. We plan to perform similar observations using the 40 cm aperture StarDICE telescope equipped with *ugrizy* filters in order to verify the grey character of the extinction. We will also assess the absolute accuracy of the correction model, as the StarDICE instrument was carefully calibrated in the laboratory and on site with a Collimated Beam Projector (T. Souverin et al. 2025).

To fully leverage the techniques explored here, ground-based telescopes conducting time-domain astronomy surveys (e.g. the Vera C. Rubin Observatory; LSST Science Collaboration 2009) could benefit from installing an infrared instrument in a nearby, independent roll-off roof building to minimize thermal fluctuations from the surrounding environment. Along with ancillary equipment such as a GNSS receiver for precise PWV retrieval and a weather station for accurate inputs to radiative transfer simulations, this setup would allow the use of data collected under variable conditions, and corrected. This is especially crucial for constructing light curves of transient objects when only a few observations are available. While absolute flux calibration remains a challenge, our proposed grey extinction correction model would facilitate transferring absolute calibration from, e.g. the CALSPEC standards (R. C. Bohlin et al. 2014) to transient objects within the same exposures.

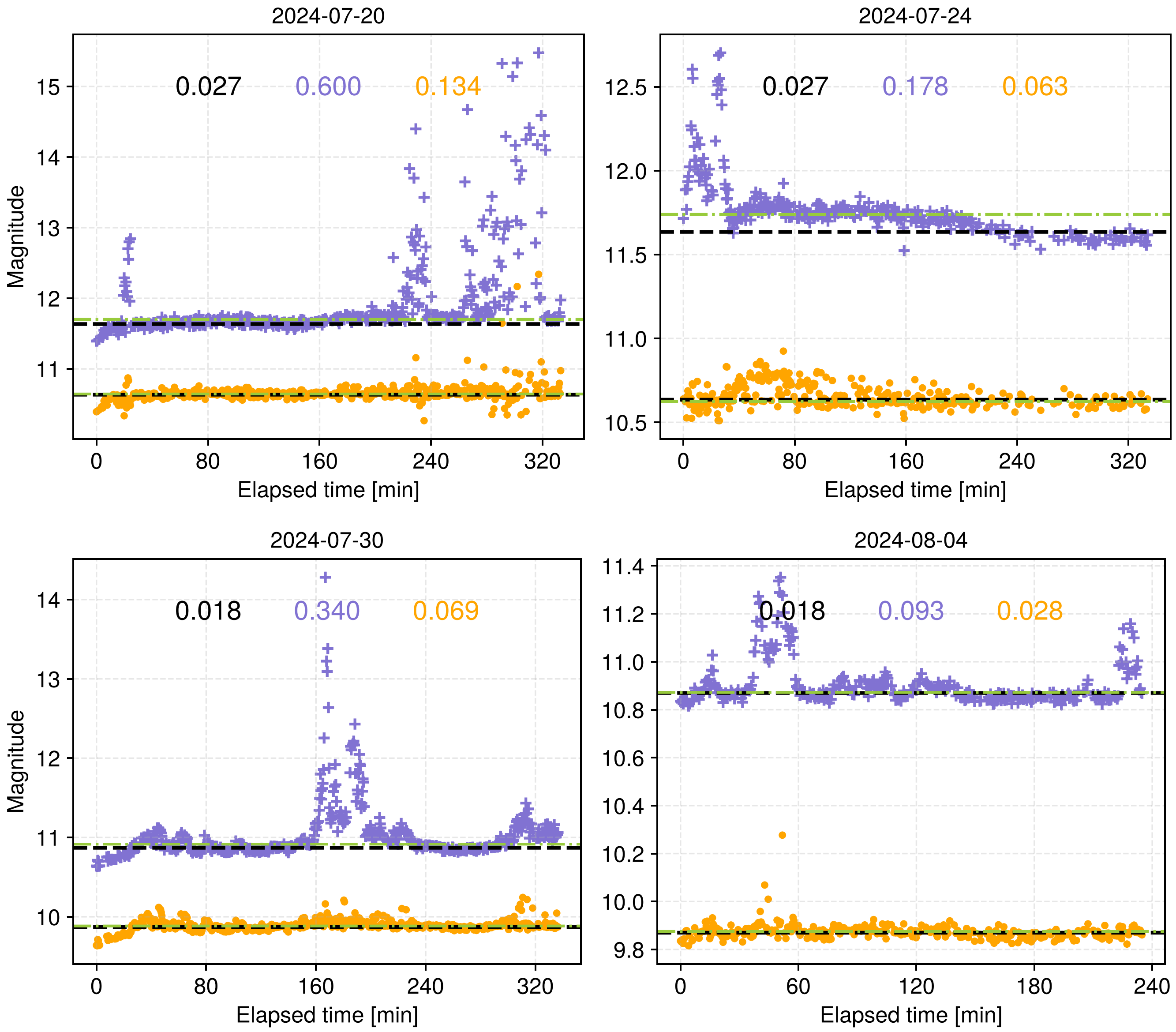

**Figure 11.** Light curves for two randomly selected stars belonging to the test catalogue of fields BD+28 4211 (top) and HD 210 955 (bottom), and two observation sequences. Reference magnitudes are indicated as black dashed lines. The median magnitudes of each sequence are depicted as green dash-dotted lines. Blue crosses are uncorrected observations. Orange circles are observations corrected by the models for grey extinction and shifted by −1.0 mag for better readability. Error bars are not shown for the same reason. The three numbers in each panel are respectively from left to right: the predicted photometric noise (black), the standard deviation of uncorrected observations (blue) and corrected observations (orange) both quadratically subtracted from the photometric noise.

## ACKNOWLEDGEMENTS

This work received support from the Programme National Cosmology et Galaxies (PNCG) of CNRS/INSU with INP and IN2P3, co-funded by CEA and CNES and from the DIM ACAV program of the Île-de-France region. The radiosonde data used in this publication were obtained as part of the Network for the Detection of Atmospheric Composition Change (NDACC) and are publicly available. This research was achieved using the POLLUX data base (https://pollux.oreme.org/) operated at LUPM (Université de Montpellier – CNRS, France) with the support of the PNPS and INSU. This work has made use of data from the European Space Agency (ESA) mission *Gaia* (https://www.cosmos.esa.int/gaia), processed by the *Gaia* Data Processing and Analysis Consortium (DPAC, https://www.cosmos.esa.int/web/gaia/dpac/consortium). Funding for the DPAC has been provided by national institutions, in particular the institutions participating in the *Gaia* Multilateral Agreement. This research has made use of the VizieR catalogue access tool, CDS, Strasbourg, France (F. Ochsenbein et al. 2000). The original description of the VizieR service was published in F. Ochsenbein et al. (2000). We thank Philippe Goloub for his efforts in establishing and maintaining the OHP_OBSERVATOIRE AERONET site, and for providing access to the data used in this work. We thank François Dolon, François Huppert and Marc Ferrari from OHP, OSU – Institut

Pythéas, UAR 3470, CNRS, Aix-Marseille Université, for their assistance in facilitating the experimental setup installation and operation at Observatoire de Haute-Provence. Some of the results in this paper have been derived using the ASTROPY, ASTROQUERY, DUSTMAPS, DUST_EXTINCTION, LMFIT, NUMPY, PHOTUTILS, SEP, and SCIPY packages. Figures in this article have been created using MATPLOTLIB. The authors thank the referee for his constructive comments that helped improve the quality and readability of the paper.

## CONFLICT OF INTEREST

Authors declare no conflict of interest.

## DATA AVAILABILITY

The data underlying this article will be shared on reasonable request to the corresponding author.

## REFERENCES

Astropy Collaboration, 2022, ApJ, 935, 167
Atmosphere U. S., 1976, US Standard Atmosphere. National Oceanic and Atmospheric Administration
Avdelidis N., Moropoulou A., 2003, Energy Build., 35, 663
Barbary K., 2016, J. Open Source Softw., 1, 58
Benirschke D., Howard S., 2017, Opt. Eng., 56, 040502
Bertin E., Arnouts S., 1996, A&AS, 117, 393
Bohlin R. C., Gordon K. D., Tremblay P. E., 2014, PASP, 126, 711
Bohlin R. C., Hubeny I., Rauch T., 2020, AJ, 160, 21
Boucaud A., 2013, Theses, Université Paris-Diderot – Paris VII, https://theses.hal.science/tel-00983440( Accessed Oct 2025)
Bradley L. et al., 2024, astropy/photutils: 1.12.0, Zenodo, https://doi.org/10.5281/zenodo.10967176
Bucholtz A., 1995, Appl. Opt., 34, 2765
Burke D. L. et al., 2010, ApJ, 720, 811
Burke D. L. et al., 2013, AJ, 147, 19
Burke D. L. et al., 2017, AJ, 155, 41
Dark Energy Survey Collaboration, 2016, MNRAS, 460, 1270
de Laverny P., Recio-Blanco A., Worley C. C., Plez B., 2012, A&A, 544, A126
DeSlover D. H., Smith W. L., Piironen P. K., Eloranta E. W., 1999, J. Atmos. Ocean. Technol., 16, 251
Emde C. et al., 2016, Geosci. Model Dev., 9, 1647
Fitzpatrick E. L., Massa D., Gordon K. D., Bohlin R., Clayton G. C., 2019, ApJ, 886, 108
Fliflet A., Manheimer W., 2006, IEEE Trans. Geosci. Remote Sens., 44, 3247
Fosu C., Hein G., Eissfeller B., 2004, Int. Arch. Photogram. Remote Sens. Spatial Inform. Sci, 35, 612
Fouesneau M. et al., 2023, A&A, 674, A28
Gaia Collaboration, 2016, A&A, 595, A1
Gaia Collaboration, 2023, A&A, 674, A1
Garrappa S. et al., 2025, A&A, 699, A50
Ginsburg A. et al., 2019, AJ, 157, 98
González-Chávez O., Cárdenas-Garcia D., Karaman S., Lizárraga M., Salas J., 2019, IEEE Trans. Instrum. Meas., 68, 4387
Gordon K. D., 2024, J. Open Source Softw., 9, 7023
Green G., 2018, J. Open Source Softw., 3, 695
Gustafsson B., Edvardsson B., Eriksson K., Jørgensen U. G., Nordlund Å., Plez B., 2008, A&A, 486, 951
Hansen J. E., Travis L. D., 1974, Space Sci. Rev., 16, 527
Hazenberg F., 2019, Theses, Sorbonne Université, https://theses.hal.science/tel-02950846( Accessed Oct 2025)
Hersbach H. et al., 2020, Quart. J. R. Meteorol. Soc., 146, 1999
Heymsfield A. J. et al., 2017, Meteorol. Monogr., 58, 2.1
Holben B. et al., 1998, Remote Sens. Environ., 66, 1
Ivezić Ž. et al., 2007, AJ, 134, 973
Jackèl D., Walter B., 1997, Comput. Graph. Forum, 16, 201
LSST Science Collaboration, 2009, preprint (arXiv:0912.0201)
Lang D., et al., 2010, AJ, 139, 1782
Larason T. C., Houston J. M., 2008, Spectroradiometric Detector Measurements::Ultraviolet, Visible, and Near-Infrared Detectors for Spectral Power. National Institute of Standards and Technology, NIST Special Publication 250-41
Lelandais L. et al., 2022, Atmos. Environ., 277, 119020
Lewis P. M., Rogers H., Schindler R. H., 2010, in McLean I. S., Ramsay S. K., Takami H., eds, Proc. SPIE Conf. Ser., Vol. 7735, Ground-based and Airborne Instrumentation for Astronomy III. SPIE, Bellingham, p. 77353C
Mahoney W., Morrison G., Matsushige G., 2012, in Peck A. B., Seaman R. L., Comeron F., eds, Proc. SPIE Conf. Ser., Vol. 8448, Observatory Operations: Strategies, Processes, and Systems IV. SPIE, Bellingham, p. 84481Y
Mayer B., Kylling A., 2005, Atmos. Chem. Phys., 5, 1855
Merienne M. F., Bekaddour A., Barbe A., 1989, in Crutzen P. J., Gerard J. C., Zander R., eds, Proc. 28th Liege International Astrophysical Colloquia, Vol. 28. p. 179
Newville M. et al., 2025, LMFIT: Non-Linear Least-Squares Minimization and Curve-Fitting for Python. Zenodo, available at: https://zenodo.org/records/15014437
Ochsenbein F., Bauer F., Marcout J., 2000, A&AS, 143, 23
Ofek E. O. et al., 2023, PASP, 135, 065001
Palacios A., Gebran M., Josselin E., Martins F., Plez B., Belmas M., Lèbre A., 2010, A&A, 516, A13
Polyanskiy M., 2024, Sci. Data, 11, 94
Reil K., Lewis P., Schindler R., Zhang Z., 2014, in Peck A.B., Benn C.R., Seaman R.L., eds, Proc. SPIE Conf. Ser., Volume 9149, Observatory Operations: Strategies, Processes, and Systems V. SPIE, Bellingham, p. 321
Riello M. et al., 2021, A&A, 649, A3
Rogalski A., Chrzanowski K., 2014, Metrol. Meas. Syst., 21, 565
Rothman L. et al., 2005, J. Quant. Sectrosc. Radiat. Transfer, 96, 139
Saastamoinen J., 1973, Bull. Geod., 107, 13
Schlafly E. F., Finkbeiner D. P., 2011, ApJ, 737, 103
Sebag J., Andrew J., Klebe D., Blatherwick R. D., 2010, in Stepp L. M., Gilmozzi R., Hall H. J., eds, Proc. SPIE Conf. Ser., Vol. 7733, Ground-based and Airborne Telescopes III. SPIE, Bellingham, p. 773348
Serrano D., Marín M. J., Núñez M., Utrillas M. P., Gandía S., Martínez-Lozano J. A., 2015, J. Atmos. Sol.-Terr. Phys., 130–131, 14
Shaw J. A., Nugent P. W., 2013, Eur. J. Phys., 34, S111
Shepard D., 1968, in Proc. 23rd ACM National Conference (ACM'68). Association for Computing Machinery, New York, p. 517
Smith W. L., Ackerman S., Revercomb H., Huang H., DeSlover D. H., Feltz W., Gumley L., Collard A., 1998, Geophys. Res. Lett., 25, 1137
Sommer K. et al., 2024, Sensors, 24, 4498
Souverin T. et al., 2024, in Bryant J. J., Motohara K., Vernet J. R. D., eds, Proc. SPIE Conf. Ser., Vol. 13096, Ground-based and Airborne Instrumentation for Astronomy X. SPIE, Bellingham, p. 130963W
Souverin T. et al., 2025, RAS Techn. Instrum., 4, rzaf010
Stamnes K., Tsay S. C., Jayaweera K., Wiscombe W., 1988, Appl. Opt., 27, 2502
Stetson P. B., 1990, PASP, 102, 932
Stubbs C. W., Tonry J. L., 2006, ApJ, 646, 1436
Stubbs C. W. et al., 2007, PASP, 119, 1163
Sugiyama J., Nishino H., Kusaka A., 2024, MNRAS, 528, 4582
Tellinghuisen J., 2020, Anal. Chem., 92, 10863
Thurairajah B., Shaw J., 2005, IEEE Trans. Geosci. Remote Sens., 43, 2000
Virtanen P. et al., 2020, Nature Methods, 17, 261
Wang S. et al., 2012, PASP, 124, 1167
Wood-Vasey W. M., Perrefort D., Baker A. D., 2022, AJ, 163, 283
York D. G. et al., 2000, AJ, 120, 1579
Young A. T., 1994, Appl. Opt., 33, 1108

Yuan X. et al., 2008, in Stepp L. M., Gilmozzi R., eds, Proc. SPIE Conf. Ser., Vol. 7012, Ground-based and Airborne Telescopes II. SPIE, Bellingham, p. 70124G

## APPENDIX A: OBSERVATION CAMPAIGN

A summary of the fields observed with the photometric imaging setup is listed in Table A1. The two fields containing BD+28 4211 and HD 210955 were ultimately chosen for in depth analysis because they (1) contained a CALSPEC standard and/or covered a wide range of airmass during the period, (2) had relatively high stellar densities and (3) were observed across thousands of exposures. Each field was tracked for several hours across a dozen of different nights with at least one night deemed 'photometric', i.e. without any cloud. The remaining fields (Polaris, HD 183030, HD 180609, and GD 153) were observed exclusively under highly variable and cloudy conditions, with poor data quality, unsuitable for further analysis. Details about each sequence of the two target fields selected for analysis are given in Tables A2 and A3. The table includes a characterization of the atmospheric conditions during each observing period; this is not a rigorous determination of the photometric quality of the data, but rather a simple summary from visual controls of all-sky camera images.

CMOS images in $r$-band were taken continuously with a constant exposure time of $t_{\mathrm{exp}} = 20$ s and fixed electronic gain of 0.6437 $\mathrm{e}^-$/ADU. This exposure time is similar to what will be used for the Rubin Observatory Simonyi Survey Telescope images to probe grey extinction structure on similar time-scales.

The time interval between successive photometric exposures is approximately 4 s, which corresponds to overhead delays from reading the image, retrieving data from ancillary instruments, writing them to the disc, and performing a thermal flat-field correction for the IR camera. We did not perform any dithering, as tracking errors and polar alignment errors were sufficient to introduce a shift of one or more pixels between successive images.

Each photometric exposure was assigned a unique integer key for later processing, allowing it to be matched with the multiple images captured by the infrared thermal camera during that exposure. The thermal camera captured images at a rate of approximately 3.5 Hz, resulting in $\sim$70 images collected during each optical exposure.

**Table A1.** Fields observed during the observation campaign. ($*$) Only these target fields were selected for the analysis. The following quantities are listed: (1) target name or identified; (2) and (3) right ascension and declination coordinates at J2000 epoch; (4) and (5) number of photometric and radiometric exposures; (6) number of nights/sequences; (7) number of remaining stellar objects after quality cuts described in Section 5.1.

| Target | RA | DEC | $N_{\mathrm{exp}}$(CMOS) | $N_{\mathrm{exp}}$(IR) | Sequences | Stars after cuts |
|---|---|---|---|---|---|---|
| Polaris | 02h31m49s | 89°15'51" | 173 | 12430 | 2 | 698 |
| HD 183030 | 17h16m56s | 89°02'16" | 834 | 63012 | 5 | 724 |
| HD 180609 | 19h12m47s | 64°10'37" | 700 | 52741 | 3 | 1235 |
| GD153 | 12h57m02s | 22°01'53" | 450 | 33832 | 1 | 330 |
| BD+28 4211* | 21h51m11s | 28°51'50" | 6996 | 479195 | 11 | 466 |
| HD 210955* | 22h13m22s | 31°22'56" | 6918 | 458225 | 12 | 476 |

**Table A2.** Summary of the observation sequences of BD+28 4211. ($*$) This night is used as the reference sequence to compute the reference magnitudes $m_s^{\mathrm{TOA,\,ref}}$. The following quantities are listed: (1) date of the sequence; (2) and (3) UTC start and end time of the sequence; (4) empirical qualitative assessment of atmospheric conditions; (5) and (6) number of photometric and radiometric exposures.

| Date [dd-mm-yyyy] | Start time [hh:mm] | End time [hh:mm] | General conditions | $N_{\mathrm{exp}}$(CMOS) | $N_{\mathrm{exp}}$(IR) |
|---|---|---|---|---|---|
| 12-07-2024 | 20:34 | 02:18 | Variables | 720 | 53782 |
| 17-07-2024 | 20:41 | 02:02 | Cloudy | 694 | 52002 |
| 18-07-2024 | 20:14 | 21:52 | Clear | 208 | 15692 |
| 19-07-2024* | 20:18 | 01:54 | Clear | 725 | 48247 |
| 20-07-2024 | 20:15 | 01:50 | Variables | 712 | 47447 |
| 21-07-2024 | 20:18 | 01:47 | Clear | 718 | 47541 |
| 23-07-2024 | 21:54 | 01:38 | Clear | 494 | 33031 |
| 24-07-2024 | 20:01 | 01:34 | Cloudy | 719 | 47970 |
| 25-07-2024 | 20:07 | 01:31 | Clear | 706 | 47033 |
| 26-07-2024 | 20:50 | 01:27 | Variables | 603 | 40336 |
| 28-07-2024 | 20:02 | 01:19 | Clear | 697 | 46114 |

**Table A3.** Summary of the observation sequences of HD 210955. (∗) This night is used as reference the sequence to compute the reference magnitudes $m_s^{\mathrm{TOA,\,ref}}$. The columns are identical to Table A2.

| Date [dd-mm-yyyy] | Start time [hh:mm] | End time [hh:mm] | General conditions | $N_{\mathrm{exp}}$(CMOS) | $N_{\mathrm{exp}}$(IR) |
|---|---|---|---|---|---|
| 29-07-2024* | 20:56 | 01:37 | Clear | 619 | 41093 |
| 30-07-2024 | 19:56 | 01:33 | Variables | 723 | 47863 |
| 31-07-2024 | 21:20 | 01:29 | Clear | 544 | 36167 |
| 04-08-2024 | 20:58 | 00:53 | Variables | 512 | 33849 |
| 05-08-2024 | 19:56 | 00:50 | Cloudy | 642 | 42692 |
| 06-08-2024 | 20:00 | 00:46 | Variables | 617 | 40693 |
| 07-08-2024 | 20:08 | 00:42 | Variables | 595 | 39222 |
| 08-08-2024 | 21:03 | 00:38 | Clear | 474 | 31722 |
| 09-08-2024 | 19:53 | 00:34 | Clear | 610 | 40066 |
| 10-08-2024 | 20:34 | 00:30 | Variables | 511 | 33587 |
| 11-08-2024 | 20:05 | 00:26 | Variables | 566 | 37327 |
| 12-08-2024 | 20:15 | 00:04 | Clear | 505 | 33944 |

## APPENDIX B: ANCILLARY DATA

Fig. B1 summarizes the evolution of atmospheric parameters throughout the observation campaign, which were used as inputs for the LIBRADTRAN atmospheric transmission and radiance simulations. Table B1 outlines the essential parameters and their retrieval methods. Here we briefly describe the pipeline to retrieve and process the ancillary data.

(i) **Barometric pressure and surface temperature** are taken from the StarDICE weather station measurements for each observation epoch (see Section 3.3).

(ii) **Precipitable water vapour (PWV)**: raw GNSS data is converted to RINEX format and then processed via the Canadian Spatial Reference System Precise Point Positioning Service (CSRS-PPP) to estimate the zenith total delay (ZTD). Zenith hydrostatic delay (ZHD) is computed using the J. Saastamoinen (1973) model with atmosphere pressure $P_{\mathrm{atm}}$ and surface temperature $T_{\mathrm{sur}}$ obtained from the weather station. Finally, zenith wet delay (ZWD = ZTD - ZHD) is converted to PWV as in J. Sugiyama et al. (2024). The visual inspection of radiance residuals between calibrated data and simulations reveals that high-frequency PWV measurements from local GNSS receivers capture subtle fluctuations more precisely than hourly interpolated satellite-based data.

(iii) **Ozone total column and ground surface albedo** are retrieved from the ERA5[18] (H. Hersbach et al. 2020) data set through the CDSAPI[19] package and by applying linear interpolation in space and time for each observation epoch.

(iv) **Aerosol properties**: aerosol optical depth and the Ångström exponent are sourced from the AERONET[20] (B. Holben et al. 1998) instrument at OHP which only provides daytime measurements. These parameters values are estimated by linear interpolation to the nighttime observation epochs.

(v) **Atmosphere vertical profiles**: temperature, humidity, and ozone profiles are derived from radiosonde data collected at OHP[21] Seasonal averages are computed with 2.5-$\sigma$ clipping from data spanning 2008–2022, and these profiles are supplemented by the US1976 standard model (U. S. Atmosphere 1976) for altitudes above 35 000 m, where atmospheric radiance and extinction contributions are minimal.

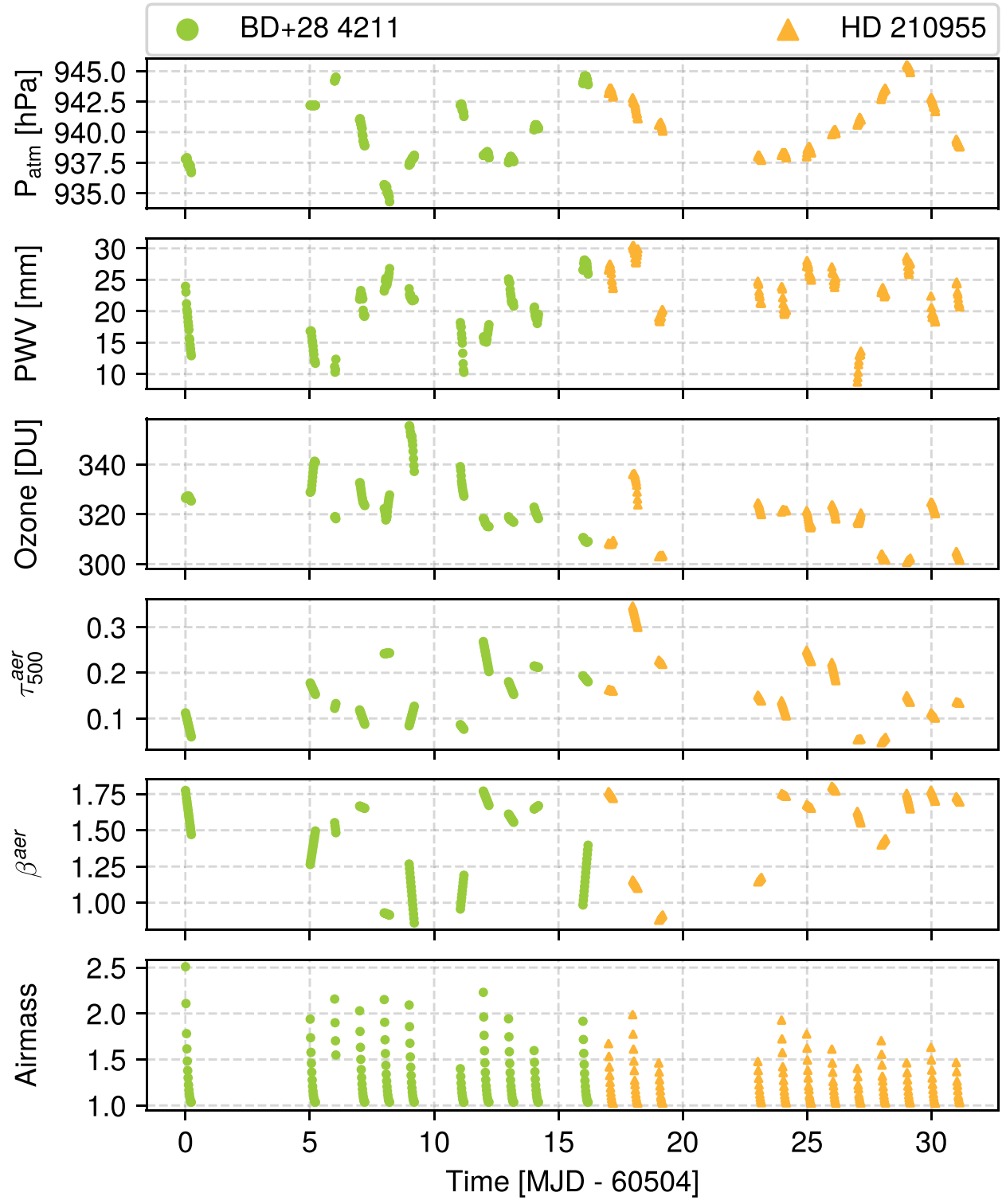

**Figure B1.** Atmospheric parameters from ancillary instrumentation and publicly available data sets given as inputs of LIBRADTRAN simulations. From the upper to the lower panel: the barometric pressure $P_{\mathrm{atm}}$, the precipitable water vapour PWV, the total ozone column $O_3$, the aerosol optical depth at 500 nm $\tau^{\mathrm{aer}}{}_{500}$, the aerosol Ångström exponent $\beta^{\mathrm{aer}}$ and the airmass at field centre. Green circles show the field of BD+28 4211 and orange triangles show the field of HD 210955. Only 1 out of 50 observations are represented to improve readability.

[18] https://cds.climate.copernicus.eu/datasets/reanalysis-era5-single-levels?tab=overview
[19] https://cds.climate.copernicus.eu/how-to-api
[20] https://aeronet.gsfc.nasa.gov/
[21] Publicly available for download from the Network for the Detection of Atmospheric Composition Change (NDACC): https://ndacc-aeris.ipsl.fr

**Table B1.** Input parameters for simulations of optical transmission and infrared radiance with LIBRADTRAN. (∗) Required to compute the cloud-free atmosphere down-welling radiance. (∗∗) Required to compute the chromatic component of optical transmission.

| Parameter | Symbol | Units | $L_{sim}^{atm}$ (∗) | $\mathcal{T}^{atm}$ (∗∗) | Retrieval method |
|---|---|---|---|---|---|
| Barometric pressure | $p^{atm}$ | hPa | ✓ | ✓ | Local weather station |
| Surface temperature | $T_{sur}$ | K | ✓ | × | Local weather station |
| Precipitable water vapour | PWV | mm, kg m$^{-2}$ | ✓ | ✓ | GNSS receiver + PPP processing |
| Ozone | $O_3$ | Dobson | ✓ | ✓ | ERA5 data base |
| Aerosol optical depth at ref. wavelength | $\tau_0^{aer}$ | – | × | ✓ | AERONET data base |
| Angstrom exponent | $\beta^{aer}$ | – | × | ✓ | AERONET data base |
| Airmass | $X$ | – | ✓ | ✓ | Astrometry + A. T. Young (1994) |
| Ground albedo | $A_g$ | – | ✓ | × | ERA5 data base |
| Vertical profiles | $\mathcal{P}^{season}$ | K, g kg$^{-1}$ | ✓ | ✓ | Seasonal average from NDACC data base |

## APPENDIX C: ATMOSPHERE EXTINCTION IN THE VISIBLE SPECTRUM

### C1 Airmass

Accurate airmass calculations based on the positions of sources are performed using the model by A. T. Young (1994). It provides an approximate formula to calculate the airmass as a function of the true zenith angle, rather than the refracted apparent zenith angle. The error in airmass calculated with this method is less than 0.001 up to $X = 6$, with a maximum of 0.0037 at the horizon (A. T. Young 1994).

### C2 Molecular extinction

Molecular Rayleigh scattering, along with absorption by oxygen and trace gases, is accurately quantified using barometric pressure data (J. E. Hansen & L. D. Travis 1974; C. W. Stubbs et al. 2007).

Atmospheric pressure varies slowly on time-scales of several hours (±1 hPa) and precise local measurements are easily done with an *in situ* weather station. By combining precise knowledge of both the airmass (Section C1) and the barometric pressure (±0.2 hPa), it is possible to provide a determination of molecular extinction with a precision of the order of a milli-magnitude on the observed stellar fluxes (C. W. Stubbs et al. 2007).

Ozone absorption can be described by a single parameter that is the integrated vertical column height from satellite-based measurements. This height is commonly expressed in Dobson Units (1 DU = 0.01 mm) and exhibits seasonal changes with minor fluctuations over several days (M. F. Merienne, A. Bekaddour & A. Barbe 1989). Consequently, linear interpolation of hourly total ozone column is sufficient.

### C3 Water vapour absorption

Water vapour absorption can be accurately simulated when the precipitable water vapour (PWV) is known. However, PWV varies rapidly over time[22], and since $H_2O$ line strengths are highly sensitive to it, water vapour absorption significantly impacts photometric measurements, particularly in the *i*, *z*, and *y* bands (see Fig. C1). Local PWV concentrations can evolve in the same way as aerosols, with variations of up to 10 per cent per hour. This requires an evolving photometric correction. Variations of the order of $\Delta PWV = 1$ mm affect the *z* and *y* band photometry of StarDICE by 5 and 25 mmag respectively. Measurements of water vapour content are made at 30 s intervals using a dedicated ancillary instrument presented in Section 3.3.

### C4 Mie scattering of aerosols

Aerosols are small particles with diameters of the same order as the incoming light, inducing Mie scattering. This phenomenon presents a chromatic dependence and is generally parametrized as a power-law model,

$$\tau^{aer}(\lambda) = \tau_0^{aer} \times \left(\frac{\lambda}{\lambda_0}\right)^{\beta^{aer}}, \quad (C1)$$

where $\tau_0^{aer}$ is the optical depth at the reference wavelength $\lambda_0$ and $\beta^{aer}$ is the Ångström exponent representing the variation of optical depth with wavelength. Assuming the aerosol cover is homogeneous, the aerosol transmission at an airmass $X$ is

$$\mathcal{T}^{aer}(\lambda, X, \tau_0^{aer}, \beta^{aer}) = e^{-\tau^{aer}(\lambda)\times X}. \quad (C2)$$

[22]After analysing the daily data collected over several months, we observed PWV fluctuations of up to 30 per cent on sub-hour time-scales at OHP.

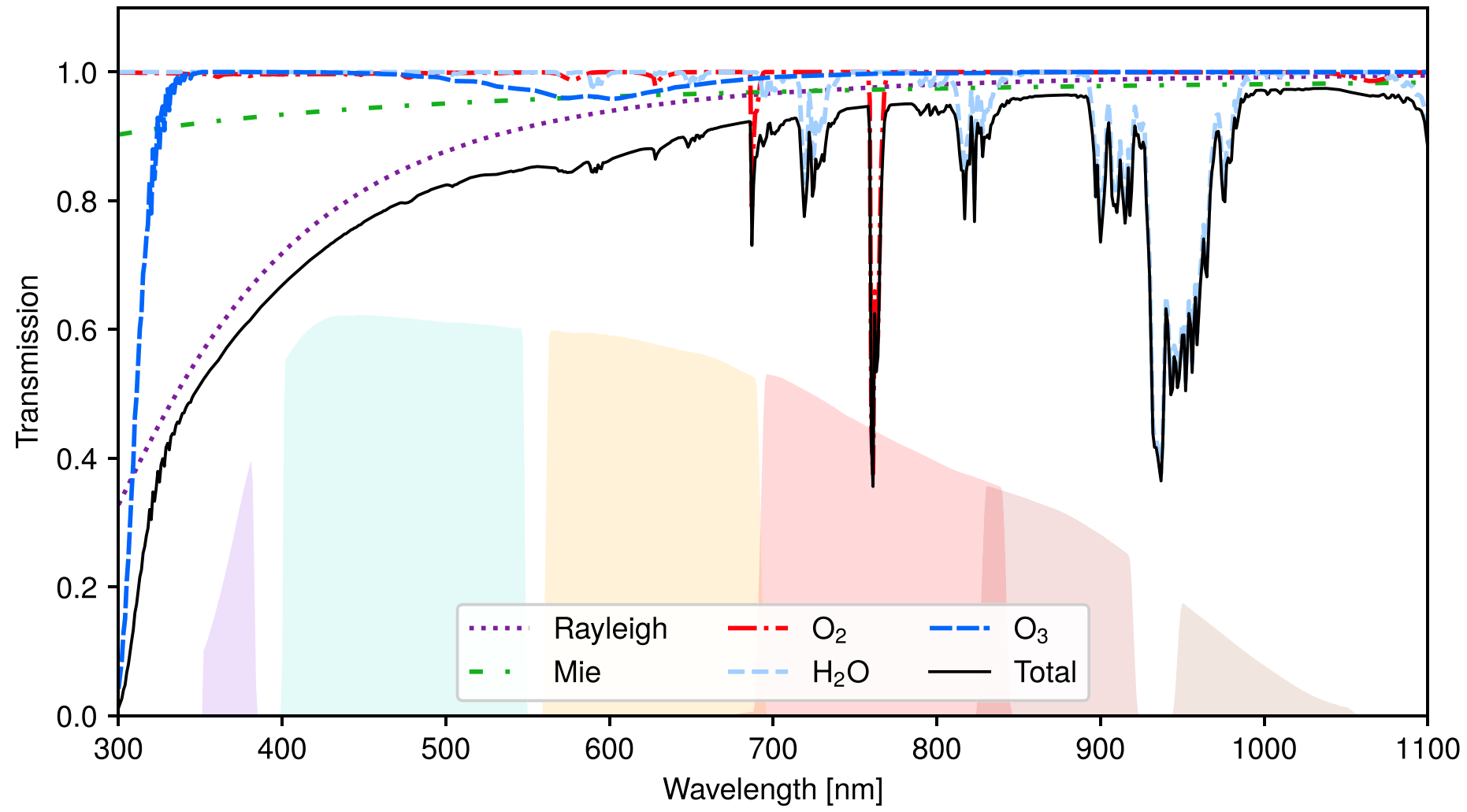

**Figure C1.** Synthetic transmission curve computed with LIBRADTRAN at airmass $X = 1$ for the OHP site location (solid black line) with the same atmosphere as Fig. 1. The Earth's atmosphere attenuates the flux from stellar sources through Rayleigh scattering (purple dotted curve), Mie scattering by aerosols with $\tau_0^{aer} = 0.05$ and $\beta^{aer} = 1.4$ (green dashed curve), molecular absorption by oxygen $O_2$ (red dashed-dotted curve), ozone $O_3 = 350$ DU (blue wide-dashed curve), and water vapour PWV = 12 mm (light blue dotted curve). The instrumental response of the StarDICE telescope *ugrizy* photometric system is also shown for reference (T. Souverin et al. 2025).

## APPENDIX D: IMPACT OF STELLAR PARAMETERS AND INSTRUMENTAL THROUGHPUT ERRORS ON CHROMATIC EXTINCTION ESTIMATION

Both the sources stellar parameters from the *Gaia* DR3 catalogue and the instrumental transmission come with uncertainties which propagate onto the synthetic spectra and the total throughput respectively. We do have an estimation on stellar parameters being ~180 K on $T_{\rm eff}$ and 0.2 dex on $\log g$ and [Fe/H] (M. Fouesneau et al. 2023). The propagated error on $\mathcal{C}_{s,i}$ for the stellar parameters uncertainty is less than 0.2 mmag for a star with $T_{\rm eff} = 5000$ K and $\log g = 0.5$ and $[{\rm Fe/H}] = 0$. It is difficult to estimate the uncertainty on instrumental transmissions for which we use typical transmissions reported by manufacturer's datasheets. The shift on $\mathcal{C}_{s,i}$ is approximately 1 mmag when adopting a conservative slope error of 30 per cent on the shape of the total instrumental throughput curve. Overall, these errors are negligible compared to the photometric measurements uncertainties which are at least one order of magnitude higher.

This paper has been typeset from a TEX/LATEX file prepared by the author.